\documentclass[aps,prl,10pt,twocolumn,superscriptaddress]{revtex4-1}

\usepackage{times}
\usepackage{changes}
\usepackage{graphicx}
\usepackage{amssymb}
\usepackage{epstopdf}
\usepackage{amsmath}
\usepackage{color}
\usepackage{hyperref}
\usepackage{bbold}
\usepackage{enumitem}
\usepackage{bm}
\usepackage{bbm}
\usepackage[acronym]{glossaries}
\usepackage{physics}
\makeatletter
\graphicspath{{Images/}}

\usepackage{hyperref}
\hypersetup{
	colorlinks=true,
	linkcolor=black,          
	citecolor=blue,           
	filecolor=magenta,        
	urlcolor=cyan
}

\makeatother

\begin{document}
	
	\title{Classical approaches to prethermal discrete time crystals in one, two, and three dimensions}
	
	\author{Andrea Pizzi}
	\affiliation{Cavendish Laboratory, University of Cambridge, Cambridge CB3 0HE, United Kingdom}
	\author{Andreas Nunnenkamp}
	\affiliation{School of Physics and Astronomy and Centre for the Mathematics and Theoretical Physics of Quantum Non-Equilibrium Systems, University of Nottingham, Nottingham, NG7 2RD, United Kingdom}
	\author{Johannes Knolle}
	\affiliation{Department of Physics, Technische Universit{\"a}t M{\"u}nchen, James-Franck-Stra{\ss}e 1, D-85748 Garching, Germany}
	\affiliation{Munich Center for Quantum Science and Technology (MCQST), 80799 Munich, Germany}
	\affiliation{Blackett Laboratory, Imperial College London, London SW7 2AZ, United Kingdom}
	
	\begin{abstract}
		We provide a comprehensive account of prethermal discrete time crystals within classical Hamiltonian dynamics, complementing and extending our recent work [\href{https://doi.org/10.1103/PhysRevLett.127.140602}{Phys. Rev. Lett. 127, 140602 (2021)}]. Considering power-law interacting spins on one-, two-, and three-dimensional hypercubic lattices, we investigate the interplay between dimensionality and interaction range in the stabilization of these non-equilibrium phases of matter that break the discrete time-translational symmetry of a periodic drive.
	\end{abstract}
	
	\maketitle
	
	\section{I. Introduction}
	Many-body systems made of a large number $N \gg 1$ of interacting elementary constituents can exhibit emergent collective phenomena, in which the whole can behave very differently from the sum of its parts \cite{anderson1972more}. Some of these phenomena can only be described by a quantum theory, whereas others can also be captured by a classical treatment. Examples of the former are quantum phase transitions like that between a Mott insulator and a superfluid (captured by a quantum Bose-Hubbard model) \cite{fisher1989boson, sachdev2007quantum}, whereas examples of the latter are finite-temperature phase transitions like the one between a ferro- and a para-magnet (captured by a classical Ising model) \cite{brush1967history}. The same distinction holds away from equilibrium: phenomena like many-body localization (MBL) necessitate a quantum theory, whereas others like thermalization can already be accounted for classically \cite{rigol2008thermalization, kardar2007statistical}.
	
	Thermalization can occur in many-body systems undergoing non-dissipative dynamics after a sudden quench (that is, change) of their Hamiltonian. According to this phenomenon, local observables \footnote{Specifically, their expectation value in the quantum case, and average over space, time, or realizations of the initial conditions in the classical case.} take under very general circumstances a steady-state value uniquely determined by the (conserved) energy of the system. Although classical and quantum mechanics can capture different flavors of the phenomenon, the rough physical intuition behind it is the same in the two descriptions: scrambling and energy redistribution among many interacting elementary constituents make the system act as a large bath for its own small sub-parts.
	
	If the physical intuition and phenomenology of thermalization are similar in classical and quantum dynamics, the two respective underlying mathematical mechanisms are not. Classical dynamics consists of $\mathcal{O}(N)$ nonlinear ordinary differential equations (ODEs) and can account for thermalization through chaos and ergodicity in phase space. Quantum dynamics consists instead of $\mathcal{O}(e^{\mathcal{O}(N)})$ linear ODEs, and captures thermalization through a peculiar spectral structure according to the so-called eigenstate thermalization hypothesis (ETH) \cite{rigol2008thermalization}.
	
	Understanding how the same phenomenon can be explained by theories with so strikingly different mathematical structures is very insightful. In this sense, the major credit of the ETH is perhaps to have resolved the seemingly paradoxical emergence of thermalization from a linear theory, rather than having discovered the phenomenon of thermalization itself \footnote{Quoting from \cite{rigol2008thermalization}, ``In generic isolated systems, non-equilibrium dynamics is expected to result in thermalization [$\dots$] However, it is not obvious what feature of many-body quantum mechanics makes quantum thermalization possible''.}. This is in line with the natural way forward of science: phenomena are generally first described/discovered within the simplest theoretical framework that can account for them, and their understanding is later refined with more and more accurate and complicated theories.
	
	Curiously enough, if this natural course from a classical to a quantum description was followed for thermalization, it was \emph{not} followed for \emph{pre}thermalization. The latter is the phenomenon for which, under a drive at high frequency $\omega$, a many-body system takes an exponentially long time $\sim e^{c\omega}$, with $c$ a constant, to heat up to an infinite temperature state
	\cite{canovi2016stroboscopic, mori2016rigorous, abanin2017effective, weidinger2017floquet, abanin2017rigorous, mallayya2019prethermalization}.
	Such a slow heating emerges from the mismatch between the large drive frequency $\omega$ and the smaller local energy scales, which provides an intuition for prethermalization both for quantum and classical systems.
	In the prethermal regime, before the ultimate heating, the system undergoes an approximate (almost energy conserving) thermalization with respect to an effective static Hamiltonian. Perhaps because of the literature on quantum Floquet engineering, prethermalization has first been discussed within a quantum framework, and only recently within a classical one \cite{rajak2018stability, mori2018floquet, rajak2019characterizations, howell2019asymptotic}.
	
	This unusual order of developments also holds for the application of prethermalization for the realization of prethermal discrete time crystals (DTCs), exotic extensions of the notion of phase of matter to the non-equilibrium domain \cite{else2017prethermal,machado2020long, luitz2020prethermalization, zhao2021random}. Prethermal DTCs are systems that break the discrete time-translational symmetry of a high-frequency drive throughout the whole prethermal regime, and represent one of the many connotations of DTCs that, after the first original proposals \cite{sacha2015modeling, khemani2016phase, else2016floquet, yao2017discrete, moessner2017equilibration}, have been put forward in various theoretical and experimental settings \cite{von2016absolute, gong2018discrete, giergiel2019discrete, matus2019fractional, gambetta2019discrete, gambetta2019classical, zhu2019dicke, kessler2019emergent, pizzi2019period, yao2020classical, pizzi2021bistability, malz2021seasonal, choi2017observation, zhang2017observation, rovny2018observation}. More specifically, prethermal DTCs respond with a periodicity multiple of that of the drive, in a way that is to some extent robust to perturbations of the Hamiltonian and initial conditions, and up to an exponentially long time $\sim e^{c\omega}$.
	
	In our accompanying work \cite{pizzi2021classicala}, we have shown that these phenomena, first studied within a quantum-mechanical approach \cite{else2017prethermal,machado2020long}, can already be captured by classical Hamiltonian dynamics (as also studied in \cite{ye2021classical}). Although it may appear as a step back in the hierarchy of available theoretical frameworks, going classical has the big advantage of lifting most of the complexity that constrains the numerics of quantum many-body systems, and that might be unnecessary for understanding prethermal DTCs, thus opening the way to the numerical simulation of systems with virtually no limits on dimension, geometry, system size, or interaction range.
	
	Indeed, in Ref.~\cite{pizzi2021classicala} we were able to study a large system of $N = 50^3$ spins in three dimensions, showing for the first time instances of prethermal DTCs in short-range interacting systems. Remarkably, the achievable large system size enabled us to study higher-order and fractional DTCs that, characterized by a period exceeding that of the drive by integer and fractional factors $n>2$, go beyond the period-doubling paradigm of MBL DTCs (in their spin 1/2 realizations) \cite{pizzi2021higher}. These dynamical phenomena are mostly elusive to exact quantum approaches on small systems, but became finally easily accessible in classical many-body dynamics \cite{pizzi2021classicala, ye2021classical}.
	
	In this paper, we build on Ref.~\cite{pizzi2021classicala} and give a detailed presentation of a generalization of the model to hypercubic lattice geometries in dimension $1,2,$ and $3$ and with tunable long-range interactions. With a comprehensive exploration of the parameter space, we aim at providing an exhaustive account of classical approaches to prethermal DTCs, and in particular at studying the interplay between dimensionality and interaction range.
	
	The remainder of the paper is organized as follows. In Section II we describe the model, including its Hamiltonian, dynamical equations, observables, and initial conditions. In Section III we present our main results. After presenting the basic phenomenology of a prethermal DTC, we compare our classical model with its quantum counterpart, and find qualitative agreement. We then showcase numerics for an array of possible parameters (specifically, drive frequency $\omega$, power-law interaction exponent $\alpha$, and strength $W$ of the noise in the initial condition) and in dimension $1,2,$ and $3$. A brief summary, discussion, and outlook on future research are presented in Section IV.
	
	Finally, let us make a remark on nomenclature. Strictly speaking, the concept of `higher-order' as introduced in Ref.~\cite{pizzi2021higher} only makes sense in relation to a system of spin $1/2$, for which the DTC period $nT$ has $n$ larger than the size $2$ of the local Hilbert space. Here, we extend this notion of `higher-order' to classical spins (for which one might say that the size of the Hilbert space is infinity) under the assumption that our findings are relevant for systems of quantum spin $1/2$, an idea that we support with numerics and discussion.
	
	\section{II. Model}
	Here, we present the model that generalizes our work on short-range interacting systems in three dimensions \cite{pizzi2021classicala} to any dimension and interaction range. This Section is organized in four subsections, discussing the systems' Hamiltonian, the respective dynamical equations, the observables of interest, and the initial condition.
	
	\subsection{II.a. Hamiltonian}
\begin{figure*}[bth]
		\begin{center}
			\includegraphics[width=\linewidth]{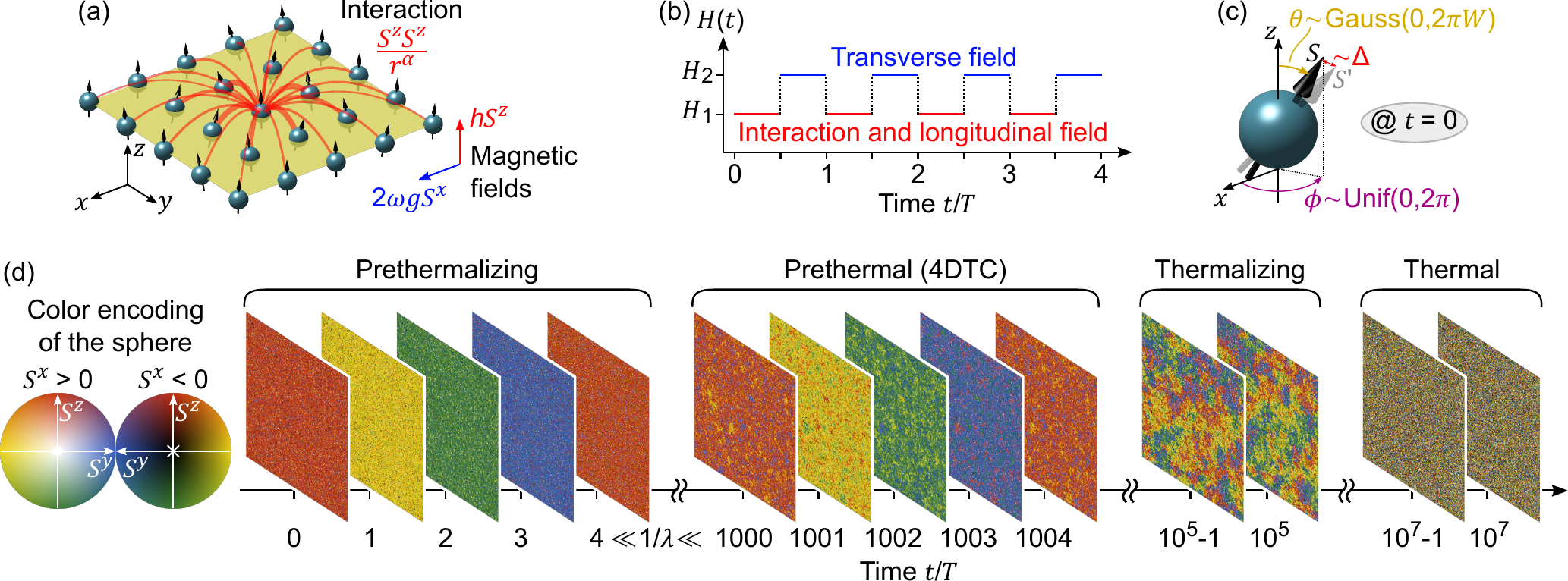}\\
		\end{center}
		\vskip -0.5cm \protect
		\caption{
			\textbf{Prethermal time crystals of (classical) spins on a hypercubic lattice}.
			(a) Schematic of the system for dimensionality $D = 2$. The spins are accomodated on a hypercubic lattice and subject to intermitting power-law interactions with characteristic exponent $\alpha$ and longitudinal and transverse fields.
			(b) The Hamiltonian ruling the system dynamics is binary: the $zz$ interaction and longitudinal $z$ field (red) are alternated with the longitudinal $x$ field (blue).
			(c) Initially ($t = 0$), each spin is moved away from the North pole by a random Gaussian polar angle $\theta$ with zero mean and standard deviation $2 \pi W$. The perturbed copy of the system $\{\bm{S}'\}$ differs from $\{\bm{S}\}$ of an amount $\sim \Delta \ll W,1$.
			(d) Phenomenology of a prethermal $4$-DTC. With a one-to-one sphere-to-color mapping, we unambiguously represent the state of a two-dimensional grid of $N = 200^2$ spins at some representative times. With a period $4$-tupled response, the spins are mostly polarized along $+\bm{z}, -\bm{y}, -\bm{z}, \bm{y}, +\bm{z}, \dots$ at times $t/T = 4k,4k+1,4k+2,4k+3,4k+4,\dots$, respectively. This is visualised by alternating predominant colorations red, yellow, green, blue, red, $\dots$. This subharmonic response holds across a long prethermal regime that extends beyond the timescale characterizing chaos, $\tau_{pth} \sim \frac{1}{\lambda}$, with $\lambda$ the Lyapunov exponent. Only at a very long time $t = 10^5 \sim \tau_{th}$ does the system order break, with the nucleation and proliferation of domains of opposite magnetizations. The infinite temperature state with random spin orientations is reached at later times $t = 10^7T \gg \tau_{th}$. Here, $g = 0.255, h = 0.1, \omega = 3.14, \alpha = \infty, R = 1,$ and $W = 0.1$.}
		\label{123PTDTCfig1}
	\end{figure*}

	Consider a hyper-cubic lattice with linear size $L$ and in dimension $D = 1,2$, or $3$. Each of the $N = L^D$ lattice sites, henceforth indexed with an index $i = 1,2,\dots, N$, hosts a classical spin $\bm{S}_i = (S_i^x, S_i^y, S_i^z)$. The spins undergo driven Hamiltonian dynamics, according to the following periodic, binary, classical Hamiltonian
	\begin{equation}
	    H(t) = 
	    \begin{cases}
	        \frac{1}{\mathcal{N}_{\alpha}} \sum\limits_{i,j=1}^N \frac{S_i^z S_j^z}{\left(r_{i,j}\right)^\alpha}
	        + h \sum_{i=1}^N S_i^z \ & \text{if} \ \text{mod}(t,T) < \frac{T}{2} \\
	        2 \omega g \sum\limits_{i=1}^N S_i^x \ & \text{if} \ \text{mod}(t,T) \ge \frac{T}{2},
	    \end{cases}
	    \label{eq123D. H}
	\end{equation}
	with
	\begin{equation}
	    \mathcal{N}_{\alpha} = \sum_{i=1}^N \frac{1}{\left(r_{i,1}\right)^\alpha}
	    \label{eq123D. Kac}
	\end{equation}
	a Kac-like normalization factor ensuring that the magnitude of the interaction term in Eq.~\eqref{eq123D. H} does not depend on $\alpha$. To avoid an unphysical self-interaction of the spins we set $r_{i,i} = \infty$, whereas for $i \neq j$ we take
	\begin{equation}
	    r_{i,j} = \sqrt{
	    \sum\limits_{\nu = x,y,z}
	    \frac{L}{\pi} \left| \tan \left[ \frac{\pi}{L}(\nu_i - \nu_j) \right] \right|^2}.
	    \label{eq123D. rij}
	\end{equation}
	For $|x_i-x_j| \ll L$, $|y_i-y_j| \ll L$, and $|z_i-z_j| \ll L$, the distance in Eq.~\eqref{eq123D. H} reduces to the familiar Euclidean one, $r_{i,j} \approx \sqrt{(x_i-x_j)^2+(y_i-y_j)^2+(z_i-z_j)^2}$. When one among $|x_i-x_j|$, $|y_i-y_j|$, and $|z_i-z_j|$ is close to $\frac{L}{2}$, instead, the tangent makes $r_{i,j}$ artificially diverge, which we expect to reduce finite-size effects, while accounting for periodic boundary conditions. A schematic of the system and drive is shown in Fig.~\ref{123PTDTCfig1}(a,b). Spherical coordinates are defined with $\bm{S}_i = \left( \sin \theta_i \cos \phi_i, \sin \theta_i \sin \phi_i, \cos \theta_i \right)$.
	
	The power law exponent $\alpha$ controls the range of the interactions. As a limit case, $\alpha = \infty$ corresponds to contact (nearest-neighbor) interactions. Henceforth, we shell consider values of $\alpha$ larger than the lattice dimensionality $D$, so that the sums in Eqs.~\eqref{eq123D. H} and \eqref{eq123D. Kac} converge in the thermodynamic limit $N \to \infty$. Indeed, even though the two divergences would compensate in Eq.~\eqref{eq123D. H} ensuring a well-defined extensive Hamiltonian \cite{khasseh2019many}, one might expect that for $\alpha \le D$ fluctuations are suppressed, rendering the physics effectively single-body. In turn, this might prevent the explicit observation of the prethermal to thermal transition, that is of prime interest for our work.
	
	\subsection{II.b. Dynamics}
	The system undergoes classical Hamiltonian dynamics. This is obtained from the Poisson brackets between the spin components and the Hamiltonian, $\dot{S}_i^{\alpha} = \left\{S_i^\alpha, H(t)\right\}$, straightforwardly computed from $\left\{ S_i^\alpha, S_j^\beta \right\} = \delta_{i,j} \epsilon_{\alpha, \beta, \gamma} S_i^\gamma$, with $\delta_{i,j}$ the Kronecker delta, $\epsilon_{\alpha, \beta, \gamma}$ the Levi-Civita anti-symmetric symbol, and $\alpha,\beta,$ and $\gamma$ in $\{x,y,z\}$.
	
	The resulting set of $3N$, coupled, nonlinear, ordinary differential equation reads
	\begin{equation}
	\frac{d \bm{S}_i}{dt} = 
	\begin{cases}
	\kappa_i \bm{z} \cross \bm{S}_i \ & \text{if} \ \text{mod}(t,T) < \frac{T}{2} \\
	2 \omega g \bm{x} \cross \bm{S}_i \ & \text{if} \ \text{mod}(t,T) \ge \frac{T}{2},
	\end{cases}
	\label{eq123D. ODE}
	\end{equation}
	where $\kappa_i$ is an effective field along $z$ accounting for both the original longitudinal field $h$ and the interaction of $\bm{S}_i$ with the other spins,
	\begin{equation}
		\kappa_i = h + \frac{1}{\mathcal{N}_{\alpha}} \sum\limits_{j=1}^N \frac{S_j^z}{\left(r_{i,j}\right)^\alpha}.
		\label{eq123D. effective kappa}
	\end{equation}
	
	Generalizing the one-dimensional short-range model considered by  Howell and collaborators \cite{howell2019asymptotic}, we can integrate the two halves of Eq.~\eqref{eq123D. ODE} getting
	\begin{equation}
	\bm{S}_i(nT+T) = 
	\begin{pmatrix}
	1 & 0 & 0 \\
	0 & c_{2,i} & -s_{2,i} \\
	0 & s_{2,i} & c_{2,i} \\
	\end{pmatrix}
	\begin{pmatrix}
	c_{1,i} & -s_{1,i} & 0 \\
	s_{1,i} & c_{1,i} & 0 \\
	0 & 0 & 1 \\
	\end{pmatrix}
	\bm{S}_i(nT)
	\label{eq123D. map}
	\end{equation}
	with $c_{1,i} = \cos \left( \kappa_i T/2\right)$, $s_{1,i} = \sin \left( \kappa_i T/2\right)$, $c_{2,i} = \cos 2 \pi g$, and $s_{2,i} = \sin 2 \pi g$. The matrix on the right in Eq.~\eqref{eq123D. map} accounts for the first half of the period and performs a rotation around the $z$ and axis under the action of the effective field $\kappa_i \bm{z}$. The left matrix performs instead the rotation around the $x$ axis due to the field $2 \omega g \bm{x}$. Iteratively applying the map in Eq.~\eqref{eq123D. map} we can efficiently evolve the system for stroboscopic times $t = 0, T, 2T, \dots$.
	
	\subsection{II.c. Observables}
	The first observable of interest is the energy averaged over one period, that is
	\begin{equation}
		H_T = \frac{1}{2\mathcal{N}_{\alpha}} \sum\limits_{i,j=1}^N \frac{S_i^z S_j^z}{\left(r_{i,j}\right)^\alpha}
		+ \sum_{i=1}^N \left(\frac{h}{2} S_i^z + \omega g S_i^x\right).
	\end{equation}
	Since the dependence on $H_T$ on $\omega$ might be disturbing when performing scaling analysis on $\omega$ itself, it is convenient to also consider a $\omega$ independent Hamiltonian. The obvious choice in this direction is that of $H(t)$ during the first half of the drive, that is
	\begin{equation}
		H_1 = \frac{1}{\mathcal{N}_{\alpha}} \sum\limits_{i,j=1}^N \frac{S_i^z S_j^z}{\left(r_{i,j}\right)^\alpha}
		+ h \sum_{i=1}^N S_i^z.
	\end{equation}
	Further, we consider the magnetization of the system along the $z$ direction
	\begin{equation}
		m = \frac{1}{N} \sum_i^N S_i^z,
		\label{eq123D. m}
	\end{equation}
	and its Fourier transform
	\begin{equation}
		\tilde{m}(\omega') = \frac{1}{M} \sum_{n = 0}^{M-1} m(nT)e^{-i \omega' nT},
		\label{eq123D. mfft}
	\end{equation}
	computed over a number of periods $M$. The choice of $M$ will stem from a tradeoff between having the desired frequency resolution ($M$ high enough) and not sampling the beyond prethermal regime ($M$ small enough). Furthermore, being interested in diagnosing chaos in the dynamics, we introduce a measure of the distance between two initially very close copies of the system, $\{\bm{S}_i\}$ and $\{\bm{S}_i'\}$ with $\bm{S}_i(0) \approx \bm{S}_i'(0)$.
	\begin{equation}
		d(t) = \sqrt{\frac{1}{N} \sum_i^N \left(\bm{S}_i(t) - \bm{S}_i'(t) \right)^2}.
		\label{eq123D. d}
	\end{equation}
	This measure, that we shell henceforth call 'decorrelator` \cite{bilitewski2018temperature, bilitewski2020classical}, directly probes the hallmark of chaos: sensitivity to the initial condition. In spherical coordinates, the copy of the system $\{\bm{S}_i'\}$ is initialized as
	\begin{align}
	\theta_{\bm{r}}'(0) & = \theta_{\bm{r}}(0) + 2\pi\Delta\delta_{\theta,\bm{r}} \\
	\phi_{\bm{r}}'(0) & = \phi_{\bm{r}}(0) + 2\pi\Delta\delta_{\phi,\bm{r}}
	\label{eq123D. S'}
	\end{align}	
	with $\delta_{\theta,\bm{r}}$ and $\delta_{\phi,\bm{r}}$ Gaussian random numbers with zero mean and standard deviation $1$. The parameter $\Delta \ll W,1$ therefore controls the initial distance between the two copies of the system $\{\bm{S}_i\}$ and $\{\bm{S}_i'\}$, see also Fig.~\ref{123PTDTCfig1}(c).
	
	The initial value of the decorrelator is small and set by the perturbation strength, $d(0) \sim \Delta$. At short times, $d$ is expected to grow exponentially, $d \sim d(0)e^{\lambda t}$, with the rate of the growth being controlled by some Lyapunov exponent $\lambda$. At very long times, $d$ is expected to take its infinite temperature value $d_\infty = \sqrt{2}$, corresponding to completely random spin orientations for the two system copies. The focus of this paper are the intermediate times, at which $1 \sim d < d_\infty$.
	
	To reduce the temporal fluctuations in the results, these global observables are possibly averaged over $R \gg 1$ independent realizations of the initial condition. Note that, anyway, we expect the fluctuations in the global observables of interest to vanish in the thermodynamic limit $N \to \infty$, even for a single realization.
	
	\subsection{II.d. Initial condition}
	The initial condition is schematically shown in Fig.~\ref{123PTDTCfig1}(c), and build as follows. We start from spins all aligned in the $z$ direction, corresponding to polar angles $\theta_i = 0$. To bring the many-body character into play, we perturb the polar angles with Gaussian noise with zero mean and standard deviation $2 \pi W$, while the azimuthal angles $\phi_i$ are taken as uniformly distributed in the entire range from $0$ to $2\pi$. In formulae, we have
	\begin{align}
	  \theta_i(0) & \sim \text{Gauss}(0,2 \pi W), \ & p(\theta) = \frac{e^{-\frac{1}{2}\left(\frac{\theta}{2 \pi W}\right)^2}}{2\pi W \sqrt{2 \pi}} \\
	  \phi_i(0) & \sim \text{Unif}(0,2\pi), \ & p(\phi) = \frac{1}{2\pi} \mathbbm{1}_{[0 \le \phi < 2\pi]}
	  \label{eq123D. IC}
	\end{align}
	with $p$ denoting the probability density function and $\mathbb{1}_{[0<\phi\le 2\pi]}$ the indicator function equal to $1$ when $0<\phi\le 2\pi$ and to $0$ otherwise. Because the perturbation has an axial symmetry with respect to the $z$ axis, for $N \to \infty$ we have $\frac{1}{N} \sum_i^{N}S_i^{x,y} = 0$. The magnetization $m = \frac{1}{N} \sum_i^{N}S_i^z$ along $z$ instead goes from $1$ for $W = 0$ to $0$ for $W = \infty$. In this sense, $W$ can be thought of as a sort of temperature of the initial condition.
	
	\begin{figure*}[bth]
		\begin{center}
			\includegraphics[width=\linewidth]{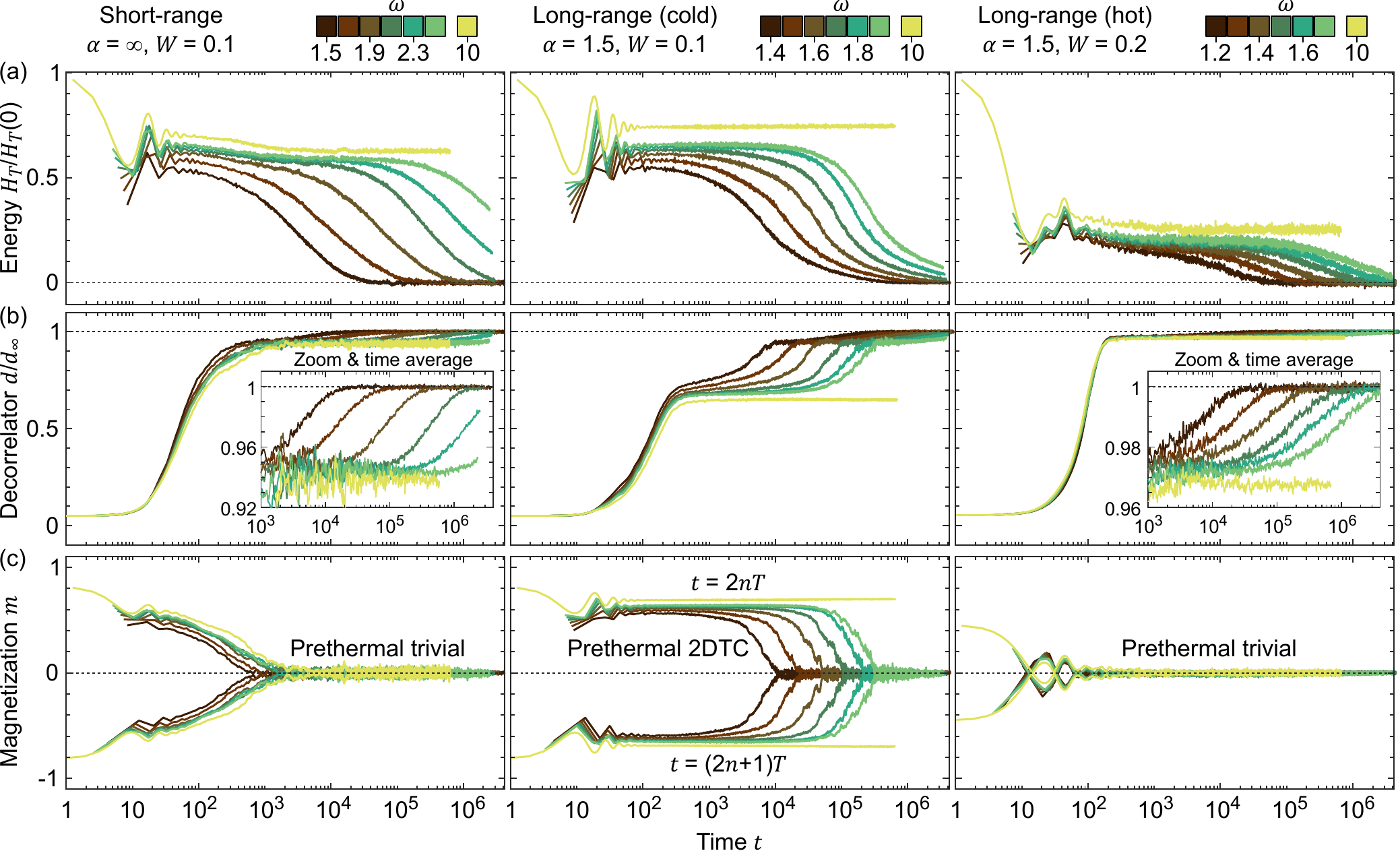}\\
		\end{center}
		\vskip -0.5cm \protect
		\caption{
			\textbf{Quantum vs classical: phenomenology of a prethermal discrete time crystal}. We show that the phenomenology of a prethermal DTC is essentially the same in classical and quantum systems by drawing a close analogy with Fig.~1 in \cite{machado2020long}. As observables, we consider (a) the energy $H_T$ averaged over one period, (b) the decorrelator $D$ measuring the distance between two initially very similar copies of the system (renormalized by its infinite temperature value $D_\infty = \sqrt{2}$), and (c) the magnetization $m = \langle s^z_i\rangle_{i,\mathrm{runs}}$. By making the association decorrelator $\leftrightarrow$ entanglement entropy, the full phenomenology of the quantum prethermal $2$-DTC described in \cite{machado2020long} is recovered: (i) for a short-range model ($\alpha = \infty$, left column) heating occurs-- over an exponentially long (in frequency) timescale - the signature of `standard' prethermalization; (ii) for a long-range model and `cold' initial condition ($\alpha = 1.5$ and $W = 0.1$, central column), prethermalization comes with the realization of a nontrivial time-crystalline (subharmonic) response of the magnetization $m$; (iii) with long-range interactions, `standard' prethermalization is recovered for a `hot' initial condition ($\alpha = 1.5$ and $W = 0.2$, right column). Here, we used $N = 100, R = 100, h = 0.1, g = 0.515, \Delta = 0.01$.}
		\label{123PTDTCfig2}
	\end{figure*}

	\begin{figure*}[bth]
		\begin{center}
			\includegraphics[width=\linewidth]{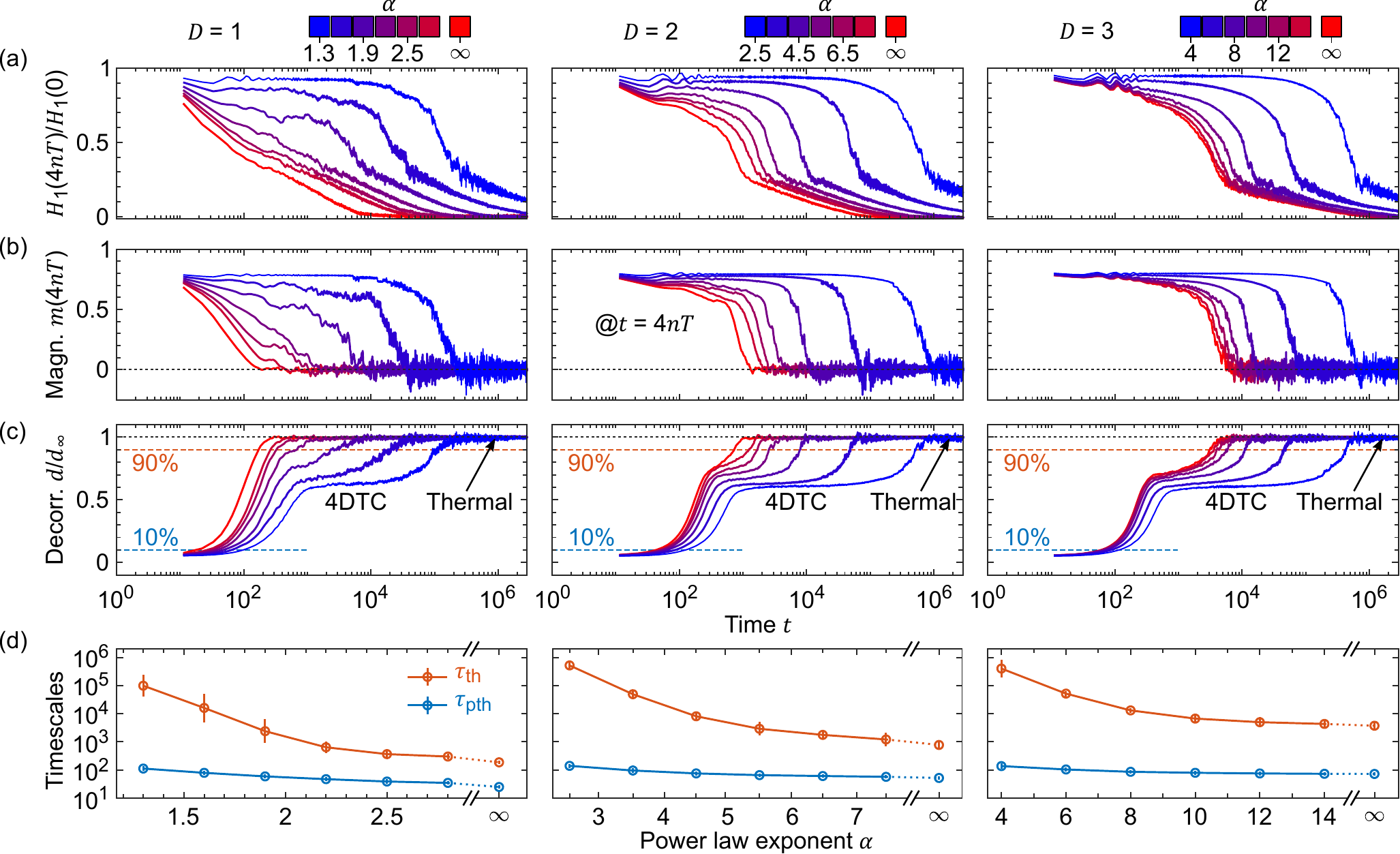}\\
		\end{center}
		\vskip -0.5cm \protect
		\caption{
			\textbf{Interplay of dimensionality and interaction range in prethermal discrete time crystals.}. We characterize the prethermal $4$-DTC by looking at (a) the energy $H_1$ (at stroboscopic times $t = 4nT$), (b) the magnetization $m = \langle s_i^z \rangle_{i, \mathrm{runs}}$, and (c) the decorrelator $d$. In one dimension (left column), a prethermal response emerges for sufficiently long-ranged interactions ($\alpha \lessapprox 2$). The thermalization as well as the prethermalization times increase with decreasing $\alpha$ (d). The phenomenology is similar in two dimensions, where the separation of timescales between prethermalization and thermalization can be well appreciated for a broader range of $\alpha \lessapprox 6$. In three dimensions, crucially, the separation of timescales extends all the way to $\alpha = \infty$. Here, we used $N = 200, 400, 343$ in $D = 1,2,$ and $3$, respectively, and $\omega = 2.2, h = 0.1, g = 0.26$ and $\Delta = 0.01$. Solid lines and errorbars are obtained as averages and standard deviations over $R = 50$ independent runs.}
		\label{123PTDTCfig3}
	\end{figure*}

	\begin{figure*}[bth]
		\begin{center}
			\includegraphics[width=\linewidth]{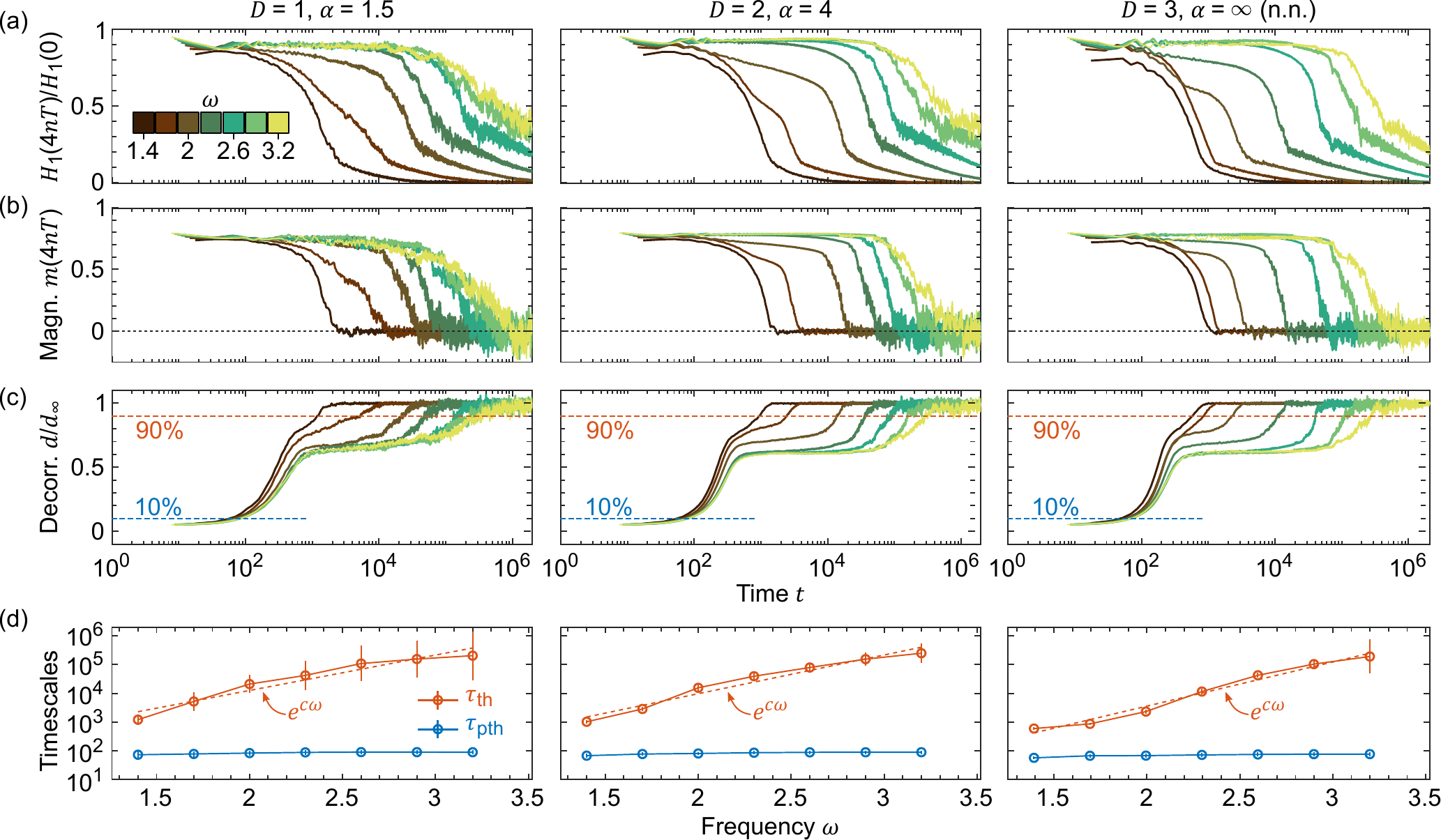}\\
		\end{center}
		\vskip -0.5cm \protect
		\caption{
			\textbf{Scaling with frequency}. (a-c) We investigate the dependency of the prethermal $4$-DTC on the frequency $\omega$ by means of the same diagnostics as in Fig.~\ref{123PTDTCfig3}. For  an interaction range $\alpha = 1.5, 4,$ and $\infty$ in $D = 1,2$ and $3$, respectively, the observed phenomenology is the same: prethermalization occurs over a timescale $\tau_{pth} \approx 1/\lambda$ that barely depends on the frequency $\omega$, whereas the full thermalization to an infinite temperature state occurs at a much later time $\tau_{th} \sim e^{c\omega}$. (d) The two timescales $\tau_{pth}$ and $\tau_{th}$ are extracted as the times at which the decorrelator crosses the $10\%$ and $90\%$ of its maximum value $D_{\infty}$, that is, the times at which its prethermal plateau can be considered to start and end. For this figure we used the same parameters as in Fig.~\ref{123PTDTCfig3}.}
		\label{123PTDTCfig4}
	\end{figure*}
	
	\begin{figure*}[bth]
		\begin{center}
			\includegraphics[width=\linewidth]{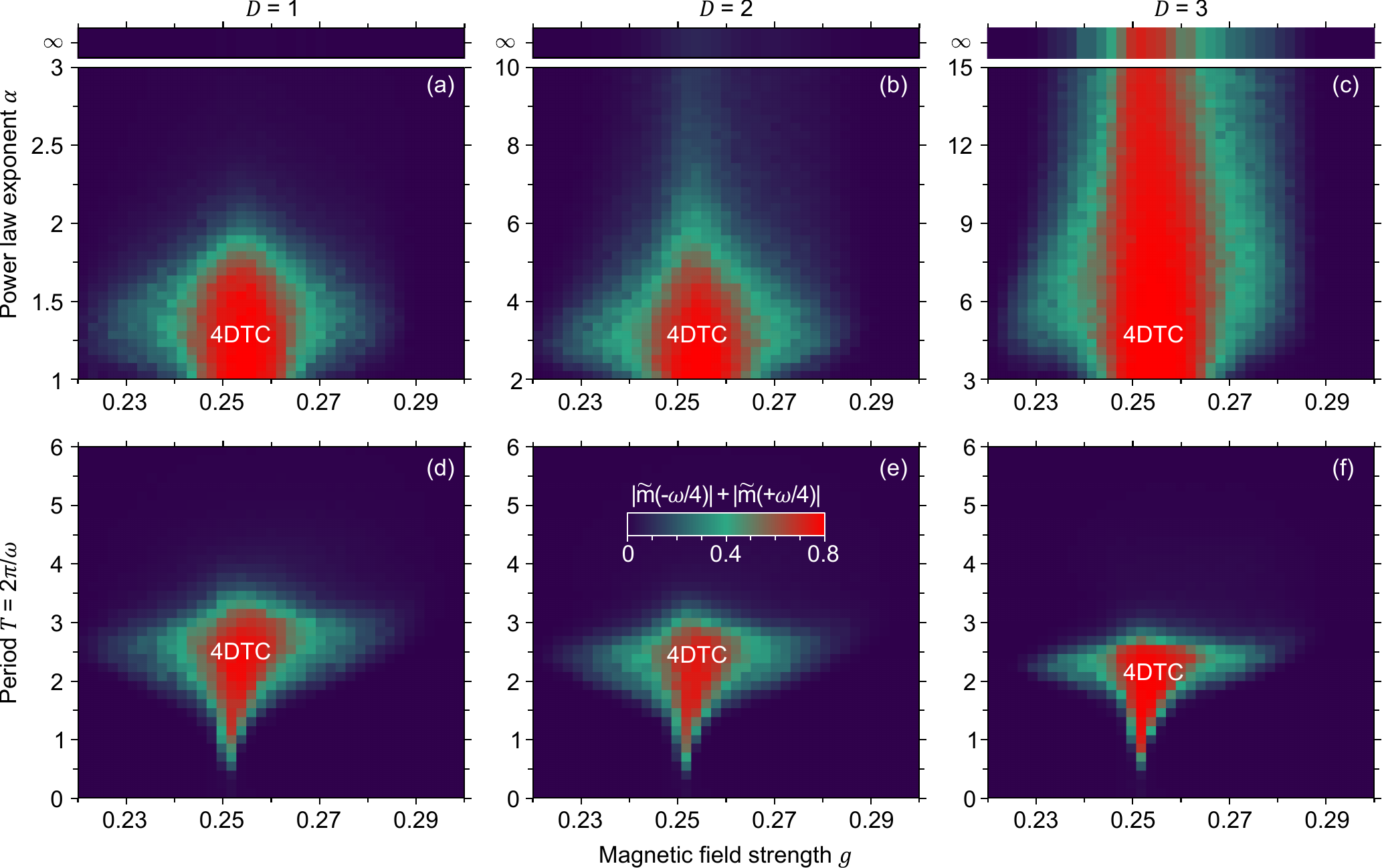}\\
		\end{center}
		\vskip -0.5cm \protect
		\caption{
			\textbf{Phase diagrams}. We investigate the stability of the prethermal $4$-DTC for dimensionality $D = 1$ (left), $2$ (center), and $3$ (right). As a diagnosis for the $4$-DTC we use the subharmonic spectral function, that is, the Fourier transform $\tilde{m}$ of the magnetization at frequency $\omega/4$, computed over the first $10^4$ periods. (a-c) Phase diagrams in the plane of the transverse field strength $g$ and power law exponent $\alpha$. In one and two dimensions (a,b) the $4$-DTC is stable for interactions that are sufficiently long-range, that is, for $\alpha < \alpha_c$, with $\alpha_c \approx 2$ and $6$ for $D = 1$ and $2$, respectively. In striking contrast, in $D = 3$ dimensions (c) the $4$-DTC extends all the way up to $\alpha = \infty$, that is, to a the short-range (nearest-neighbor) limit. (d-e) Phase diagrams in the plane of transverse field $g$ and drive period $T = 2\pi/\omega$. The phase diagram looks qualitatively the same in all dimensions: the frequency should be large enough for the $4$-DTC to be stable (in a prethermal fashion), but the larger the frequency and the smaller the range of $g$ over which the $4$-DTC is stable. Here, we used $N = 150, 144,$ and $216$ in $D = 1,2,$ and $3$, respectively, $T = 2.5$ in (a-c), $\alpha = 1.5, 4,$ and $\inf$ in (d-f), respectively, $R = 50$, $h = 0.1$, and $W = 0.1$.}
		\label{123PTDTCfig5}
	\end{figure*}

	\begin{figure}[bth]
		\begin{center}
			\includegraphics[width=\linewidth]{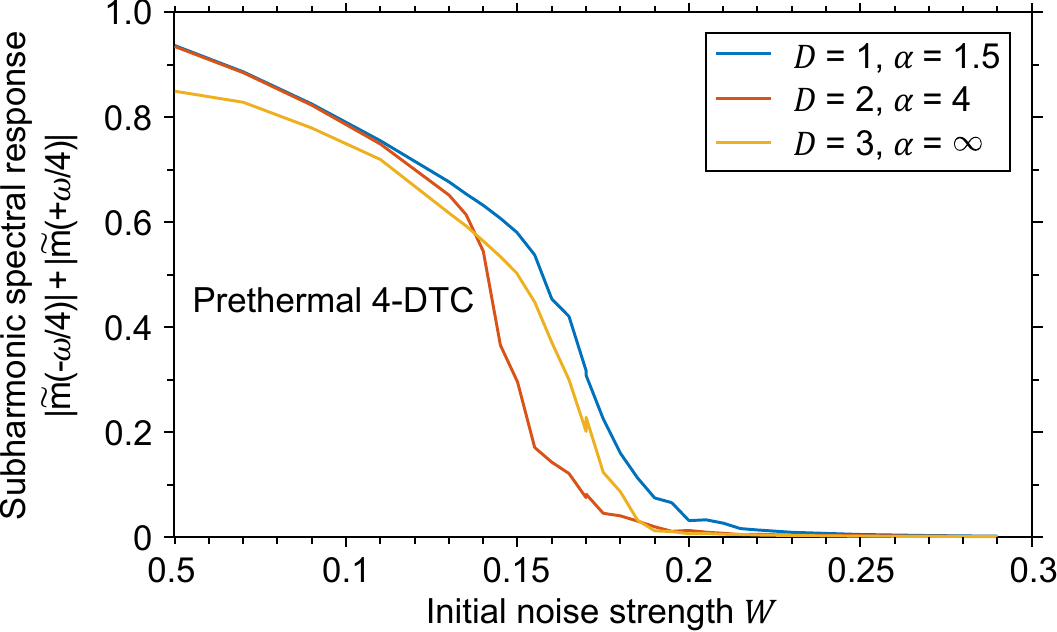}\\
		\end{center}
		\vskip -0.5cm \protect
		\caption{
			\textbf{Stability against perturbation of the initial condition}. We investigate the stability of the prethermal $4$-DTC with respect to perturbations of the initial condition, for dimensionality $D = 1, 2,$ and $3$ and for power law exponent $\alpha = 1.5, 4$, and $\infty$, respectively. As in Fig.~\ref{123PTDTCfig5}, we compute the subharmonic spectral function at frequency $\omega/4$ over the first $5000$ periods, and use it as an order parameter for the $4$-DTC. We observe a phase transition at a certain finite critical value of the noise strength $W_c \sim 0.18$, above which the prethermal DTC disappears in favour of trivial thermalization. Note, the observed fluctuations are a finite-size effect. Here, we used $N = 2000, 2500,$ and $8000$ in $D = 1,2,$ and $3$, respectively, whereas $T = 2.5$, $R = 20$, $g = 0.25$, and $h = 0.1$.}
		\label{123PTDTCfig6}
	\end{figure}

	\section{III. Results}
	This section is organized as follows. First, in subsection III.a.~we present in qualitative terms the key dynamical behavior of interest: the prethermal DTC. Second, in subsection III.b.~we focus on one-dimensional systems to perform a close comparison between the phenomenology of a prethermal DTCs obtained in the classical and quantum domains, show that the two are remarkably similar, and suggest that prethermal DTCs can be studied regardless of quantum fluctuations. With the latter idea in mind, in subsection III.c.~we present a comprehensive study of prethermal DTCs for various interaction ranges and in dimension $1$, $2$, and $3$.
	
	\subsection{III.a. Prethermal discrete time crystals}
	The phenomenology of a prethermal DTC is illustrated in Fig.~\ref{123PTDTCfig1}(d) for a single realization of a two-dimensional lattice of spins with short-range nearest-neighbor interactions ($\alpha = \infty, R = 1$). To unambiguously visualise the state of all the spins at a given time, we perform a one-to-one mapping between the unit sphere and a colormap, so that every possible spin orientation corresponds to a unique colour In particular, the principal spin orientations $+\bm{x}, -\bm{x}, +\bm{y}, -\bm{y}, +\bm{z},$ and $-\bm{z}$ correspond to the colours white, black, blue, yellow, red, and green, respectively. For $g = 0.255 \approx 1/4$, the spins are rotated at every period of an angle $\approx \pi/2$. At short times $t = 0,T,2T,3T,4T,\dots \ll \tau_{pth}$, this intuitively results in a sequence of spin orientations $\approx +\bm{z}, -\bm{y}, -\bm{z}, \bm{y}, +\bm{z}, \dots$ resulting in predominant red, yellow, green, blue, red, $\dots$ coloration, respectively. What is more remarkable is that this period 4-tupled subharmonic response persists for very long times $\tau_{pth} \ll t \ll \tau_{th}$: a prethermal $4$-DTC. The times $t = 1000T, 1001T, 1002T, 1003T,$ and $1004T$ are shown as representative for the prethermal regime. At a longer time $t/T = 10^5 \approx \tau_{th}$, the spins start depolarising with the formation of domains with opposite magnetizations, giving rise to a speckled coloration. At very long times $t = 10^7 \gg \tau_{th}$, the system has reached the infinite temperature state in which each spin has a completely random orientation.
	
	More generally, a prethermal $n$-DTC is characterised by two well-separated prethermalization and thermalization timescales, $\tau_{pth} \sim 1/\lambda$ and $\tau_{th} \sim e^{c\omega}$, respectively. For $t \ll \tau_{pth}$, the system exhibits a subharmonic response but has not yet equilibrated to the effective Hamiltonian $H_{\mathrm{eff}}$ that governs the dynamics at stroboscopic times $t = T, nT, 2nT, \dots$. This is witnessed by a $d \ll 1$, meaning that the sensitivity to initial conditions has not been expressed yet. For $\tau_{pth} \ll t \ll \tau_{th}$, the system at stroboscopic times $t = nkT$ has equilibrated to the effective Hamiltonian $H_{\mathrm{eff}}$ while the dynamics has remained subharmonic with periodicity $nT$, the distinctive feature of the prethermal $n$-DTC. Sensitivity to initial conditions has now fully come into play ($d \sim 1$), and yet the system remains correlated with its initial condition ($d < d_\infty$) thanks to the time-translational symmetry breaking. At later times $t \gg \tau_{th}$, the system has ultimately reached the infinite temperature state, in which the spins have completely random orientations and memory of the initial condition have been completely lost ($d = d_\infty$).
	
	We emphasize that a key element for a prethermal DTC is that the separation between the timescales $\tau_{pth}$ and $\tau_{th}$ can be increased by simply increasing $\omega$. We also note that similar arguments hold in the case of fractional $n$-DTCs with fractional $n$.

	\subsection{III.b. One-dimensional chain: quantum vs classical}
	In this subsection we wish to answer the question to what extent prethermal DTCs are due to quantum fluctuations. To this end, we consider the phenomenon that is the most accessible to numerics: a $2$-DTC in dimension $D = 1$. In one dimension, a model analogous to that in Eq.~\eqref{eq123D. H} but for quantum spins $1/2$ and focussing on $2$-DTCs has indeed been analysed both numerically and analytically in Ref.~\cite{machado2020long}. In that work, the key phenomenology of the quantum prethermal $2$-DTC is perhaps best exemplified in Fig.~1. Thus, we build in Fig.~\ref{123PTDTCfig2} here the classical analogue of Fig.~1 in Ref.~\cite{machado2020long}. We show that the main qualitative features of the prethermal $2$-DTC almost do not change when going classical, suggesting that these non-equilibrium phenomena should be thought as robust to quantum fluctuations, rather than dependent on them.
	
	The core idea for the investigation of prethermalization and prethermal phases of matter is, almost by definition, to check how the dynamics of observables changes as a function of the drive frequency $\omega$. In Fig.~\ref{123PTDTCfig2}, we perform this analysis for the average energy $H_T$, magnetization $m$, and decorrelator $d$, in the top, middle, and bottom rows, respectively. One might think of the latter in analogy to the entanglement entropy considered in Ref.~\cite{machado2020long}, although entanglement is of course a quantum concept with no strict classical counterpart. To target the $2$-DTC, we set $g = 0.515 \approx 1/2$. Furthermore, in each column we consider three different combinations of interaction range (controlled by $\alpha$) and temperature of the initial condition (controlled by $W$). The scenarios of interest are the following:
	
	(i) \textit{Short-range} --
	For nearest-neighbor interactions in one dimension ($\alpha = \infty, D = 1$) we observe standard prethermalization. This is observed in the scaling with $\omega$ of the average energy $H_T$, that takes a time $\tau_{th} \sim e^{c\omega}$ to decay to its infinite temperature value $0$. Prethermalization also leaves its signature in the decorrelator $d$, that plateaus to a finite temperature value $\sim 0.94 \% d_\infty$ before reaching the infinite temperature value $d_\infty$ at times $t \sim \tau_{th}$ (to see the plateau more clearly, in the inset we show a close up of the data and average them over a moving time window that is short compared to the timescales of interest). This prethermalization is however of standard type, meaning that it does not realize a nontrivial prethermal DTC. Indeed, the period-doubled subharmonic response of the magnetization $m$ has a short lifetime that does not scale with $\omega$, as the thermalization time does. Rather, $m$ reaches the infinite temperature value $0$ before the infinite temperature state is reached, and discrete time-translational symmetry is not broken, not even in a prethermal fashion. As long as the interactions are short-ranged, this fate of trivial prethermalization is true regardless of the temperature of the initial condition, including the considered low temperature ($W = 0.1$).
	
	(ii) \textit{Long-range, cold} --
	For long-range interactions ($\alpha = 1.5$) and a `cold' initial condition ($W = 0.1$) we observe a prethermal $2$-DTC. Prethermalization is again diagnosed by the energy $H_T$ needing a time $\tau_{th} \sim e^{c\omega}$ to decay to its infinite temperature value $0$, as well as by the decorrelator $d$ correspondingly plateauing at a finite-temperature value $\sim 0.65 \% d_\infty$. In contrast to (i), however, prethermalization now comes along with a subharmonic response of the magnetization $m$. Discrete time-translational symmetry is broken for exponentially long (in $\omega$) times: a prethermal $2$-DTC.
	
	(iii) \textit{Long-range, hot} --
	For long-range interactions ($\alpha = 1.5$) and a `hot' initial condition ($W = 0.2$) we observe standard prethermalization. Indeed, the increase in temperature in the initial condition is sufficient to re-establish the discrete time-translational symmetry already at times $\sim 10^2 \ll \tau_{th}$, in a way similar to (i) and despite the long-range interaction. Indeed, a necessary condition for a prethermal DTC in the absence of disorder is that the temperature of the initial condition is low enough \cite{machado2020long}, which is not the case here.
	
	Making the association decorrelator $\leftrightarrow$ entanglement entropy, we have therefore shown that the phenomenology of a prethermal $2$-DTC barely changes when going from quantum to classical. This leads us to conjecture that the core underlying physics of prethermal DTCs is not `genuinely quantum', in analogy with many finite-temperature phase transitions in equilibrium statistical mechanics, that are captured by classical physics while describing systems that are intrinsically quantum (e.g., the magnetic transition of the Heisenberg model). In fact, this analogy is more than an evocative speculation, because of the strong one-to-one correspondence between prethermal DTCs and equilibrium finite-temperature transitions outlined in Refs.~\cite{else2017prethermal, machado2020long}.
	
	If it is true that the essence of prethermal DTCs is captured by classical mechanics, the way forward is then clear: use the Hamiltonian dynamics in Eq.~\eqref{eq123D. map} to study scenarios that are numerically hardly accessible with a many-body quantum-mechanical approach. Let us therefore identify the most ambitious setting for the application of this paradigm. On the one hand, in Ref.~\cite{pizzi2021higher} we showed that, in contrast to $2$-DTCs, quantum higher-order $n$-DTCs with $n>2$ are subject to much stringent finite-size constraints, as their signatures emerge only for system sizes exceeding by a factor $\sim 2$ those within the reach of exact diagonalization techniques. On the other hand, a major open challenge for quantum many-body systems is higher dimensionality $D > 1$, for which finite-size effects become even more severe. This sets the goal for the following subsection: the study of higher-order DTCs in dimension $D = 1,2,3$.

	\subsection{III.c. Interplay of interaction range and dimensionality}
	We now study the interplay between dimensionality $D$ and interaction range (controlled by $\alpha$) in the stabilization of prethermal DTCs. With focus on the higher-order $4$-DTC obtained for $g \approx 1/4$, we showcase a comprehensive exploration of the parameter space.
	
	To begin with, in Fig.~\ref{123PTDTCfig3} we investigate the effect of the interaction range by varying the power-law exponent $\alpha$ for a fixed drive frequency $\omega = 2.2$. As observables of interest, we focus on the Hamiltonian $H_1$ (first row), magnetization $m$ (second row), and decorrelator $d$ (third row). The first two are measured at stroboscopic times $t = 4kT$ according to the periodicity of the $4$-DTC (note, this is in contrast to Fig.~\ref{123PTDTCfig2}, where stroboscopic times $t = kT$ have been considered instead). In the fourth, bottom row we instead plot the times $\tau_{th}$ and $\tau_{pth}$, defined as the times at which the decorrelator $d$ crosses the values $10\% d_\infty$ and $90\% d_\infty$.
	In dimension $D = 1$ (left column), we observe that the separation between these two timescales is more prominent for $\alpha \lessapprox 2$, for which the prethermal plateau of the decorrelator at value $\sim 60\% d_\infty$ signalling a $4$-DTC starts to emerge, accompanied by a plateau of the stroboscopic $m(4kT)$. This is in agreement with the equilibrium phase transition at $\alpha = 2$, that we expect to be dual to the prethermal DTC \cite{else2017prethermal, machado2020long}. In dimensions $D = 2$ and $3$ (mid and right columns, respectively), we again observe that a longer-range interaction (that is, smaller $\alpha$) facilitates a prethermal time-crystalline response. In contrast to the $D = 1$ case, however, this persists all the way to $\alpha = \infty$, that is for a short-range (nearest-neighbor) interaction as, again, one would expect from the duality with equilibrium. The prethermal $4$-DTC for $\alpha = \infty$ is actually not fully appreciated here for $D = 2$, but this is just because the considered frequency is not large enough, and a clearer time-crystalline response appears for the parameters considered in Fig.~\ref{123PTDTCfig3}(d).
	
	For the same set of observables, we then move on to consider in Fig.~\ref{123PTDTCfig4} the scaling with $\omega$ for a fixed interaction range $\alpha$. For the considered $\alpha = 1.5, 4$, and $\infty$ in $D = 1,2$, and $3$, respectively, we observe the definiting features of a prethermal $4$-DTC. The energy $H_1$ does not decay to its infinite temperature value $0$ for a very long prethermal regime (a), during which the magnetization exhibits a period $4$-tupled subharmonic dynamics (b), and the decorrelator $d$ plateaus at a finite temperature value $\sim 60 \% d_\infty$. The increase of the separation between the timescales $\tau_{pth}$ and $\tau_{th}$, within which the prethermal $4$-DTC lives, is fully appreciated in the bottom row of Fig.~\ref{123PTDTCfig4}(d).
	
	In Fig.~\ref{123PTDTCfig5} we use the subharmonic spectral response $|\tilde{m}(-\omega/4)|+|\tilde{m}(+\omega/4)|$ at subharmonic frequency $\omega' = \omega/4$ as an order parameter to sketch some representative phase diagrams for the $4$-DTC. Again, dimensionalities $D = 1,2$, and $3$ are considered in the left, mid, and right columns. In the top row we explore the plane of the magnetic field strength $g$ and the power-law exponent $\alpha$. For $D = 1$, the stability region of the $4$-DTC embraces $g = 1/4$ and extends up to $\alpha \approx 2$, that we have already noted as the likely upper critical value in agreement with the analogue equilibrium transition. To be more precise, the critical $\alpha$ is likely slightly smaller than $2$, because the initial condition is at finite (rather than $0$) temperature, being $W = 0.1$. For $D = 2$ and $3$ we expect the prethermal $4$-DTC to be stable all the way to $\alpha = \infty$, that is for short-range (nearest-neighbor) interactions. As in Fig.~\ref{123PTDTCfig3}, the persistence of the $4$-DTC to $\alpha = \infty$ for $D = 2$ is actually not easily appreciated just because the considered frequency $\omega = 2.51$ is too low. In the bottom row we consider instead the plane of magnetic field strength $g$ and drive period $T = 2\pi/\omega$. Again, the stability region of the $4$-DTC develops around $g = 1/4$. For too large periods $T$ (too small frequency $\omega$), the subharmonic response lasts for a time much shorter than the time window used to compute the Fourier transform $\tilde{m}$ (first $10^4$ periods of the drive), and the order parameter does not detect the $4$-DTC. This changes for $T \lessapprox 3$, corresponding to a high-enough frequency for the prethermal regime to extend over a range larger than that used to compute $\tilde{m}$, thus allowing the emergence and detection of a prethermal $4$-DTC. For increasing large frequencies ($T \to 0$), we observe that the region of stability of the $4$-DTC decreases, while remaining centred around $g = 1/4$ (the final disappearance of the $4$-DTC for $T \to 0$ is an artefact of the relatively poor resolution in $g$).
	
	Finally, in Fig.~\ref{123PTDTCfig6} we again use the subharmonic spectral response $|\tilde{m}(-\omega/4)|+|\tilde{m}(+\omega/4)|$ at subharmonic frequency $\omega' = \omega/4$ as an order parameter to verify the stability of the $4$-DTC against perturbations of the initial condition. To effectively change the temperature (that is, energy) of the initial condition, we vary the strength of the initial noise $W$, see Eq.~\eqref{eq123D. IC}. As expected, we find a finite critical $W_c > 0$  above which the subharmonic response disappears, which confirms the expectations on the relation between prethermal DTCs and equilibrium finite-temperature transitions \cite{else2017prethermal, machado2020long}.
	
	We note that, for both Fig.~\ref{123PTDTCfig5} and Fig.~\ref{123PTDTCfig6}, a more careful analysis should have actually focussed on the scaling properties of the subharmonic response: for each point in the parameter space, one should repeat the simulations at various frequencies $\omega$ and fit the resulting thermalization times $\tau_{th} \sim e^{c\omega}$ to extract the scaling coefficient $c$, which should then itself be used as an order parameter. This analysis would however require some intensive numerics that goes beyond the scopes of this work. As well, we emphasize that the results shown here are not an artefact of the finite system size, as can be verified with simple scaling analysis \cite{pizzi2021classicala}.
	
	\section{IV. Discussion and conclusion}
	
	By simulating the many-body Hamiltonian (non-dissipative) dynamics of interacting classical spins on hypercubic lattices, we provided a comprehensive account of prethermal DTCs in dimension $1, 2,$ and $3$. By varying the exponent $\alpha$ of the power-law coupling, we investigated the interplay of interaction range and dimensionality. We found that the duration of the prethermal time-crystalline response increases with frequency as $\sim e^{c\omega}$ but also with interaction range. We provided numerical evidence that prethermal DTCs are possible for $\alpha \lessapprox 2$ in dimension $1$, and for any $\alpha$ in dimension $2$ and $3$, including the short-range limit $\alpha = \infty$. Our work puts forward classical Hamiltonian dynamics as a prime tool for the numerical investigation of prethermalization-related phenomena, with virtually no constraint on the system's geometry, size, and underlying Hamiltonian, and therefore of direct applicability to experiments.
	
	The lesson that a rich structure of eigenstates in an exponentially large Hilbert space can account for complex scrambling phenomena even in a linear quantum theory had first been appreciated in the context of thermalization with the ETH \cite{rigol2008thermalization}, but naturally extended to prethermalization and prethermal phases of matter \cite{else2017prethermal, machado2020long}. Having learnt this important lesson, our work now suggests that, to describe many key aspects of these phenomena, we can take a step back and adopt the much simpler classical models. Indeed, the comparison between Fig.~\ref{123PTDTCfig2} here and Fig.~1 in Ref.~\cite{machado2020long}, together with physical intuition, suggests that the phenomena studied here should be thought of as robust to quantum fluctuations, rather than dependent on them.
	
	For this reason, we expect that the higher-order prethermal DTCs described here should be observable in a wide range of experimental platforms, both inherently quantum and not. In the first group are the setups already adopted to investigate time-crystalline behaviors, such as nitrogen–vacancy (NV) spin impurities in diamond \cite{choi2017observation}, trapped atomic ions \cite{zhang2017observation}, or $^{31}$P nuclei in ammonium dihydrogen phosphate (ADP) \cite{rovny2018observation}, whereas the second group should be relevant in the context of spintronics and magnonics \cite{grundler2002spintronics, kruglyak2010magnonics}.
	
	We would like to make a remark on nomenclature. To convey the idea that both classical and quantum models can be used to shed light on the physical phenomena of prethermalization and prethermal DTCs, irrespective of their ``true'' (classical or quantum) nature, we prefer to talk of classical and quantum approaches to the prethermal DTCs, rather than classical and quantum prethermal DTCs
	\footnote{Theories and models can be labelled as classical or quantum, but if both classical and quantum theories can shed light on prethermal DTCs, then a classification of prethermal DTCs as classical or quantum is perhaps not needed.}.
	For this reason, here we did not use the word `classical' as an adjective for the phenomenon of prethermalization, including in the title, where we used brackets.
	
	As a brief outlook for future research, an insightful question regards the assessment of the exact functional form of $\tau_{th}$. We note that, numerically, this is a hard enterprise because it requires probing thermalization over a broad range of frequencies $\omega \gg 4$, when the $\tau_{th} \sim 10^6 T$ for $\omega = 3.5$ is already remarkably long and computationally demanding, see Figs.~\ref{123PTDTCfig3} and \ref{123PTDTCfig4}. Nonetheless, even if a precise characterization of $\tau_{th}$ is interesting on its own, we emphasize that what really matters here is the existence of a clear separation of timescales ($\tau_{pth}$ and $\tau_{th}$) leaving time for the emergence of prethermal DTCs. It would then be desirable to clarify to what extent quantum fluctuations affect our findings, as to some extent addressed in one dimension \cite{machado2020long, pizzi2021higher}. An important open question regards the existence of genuinely quantum (that is, with no classical counterpart) prethermal phases of matter.
	Finally, on a more open end, future research should aim at applying classical Hamiltonian dynamics to study other dynamical phenomena, both related to prethermalization and beyond it.
	
	\textbf{Acknowledgements.}
	We thank J.~Garrahan, A.~Lazarides, D.~Malz, S.~Roy, and H.~Zhao for interesting discussions on related work. We acknowledge support from the Imperial-TUM flagship partnership. A.~P.~acknowledges support from the Royal Society and hospitality at TUM. A.~N.~holds a University Research Fellowship from the Royal Society.


\begin{thebibliography}{51}%
		\makeatletter
		\providecommand \@ifxundefined [1]{%
			\@ifx{#1\undefined}
		}%
		\providecommand \@ifnum [1]{%
			\ifnum #1\expandafter \@firstoftwo
			\else \expandafter \@secondoftwo
			\fi
		}%
		\providecommand \@ifx [1]{%
			\ifx #1\expandafter \@firstoftwo
			\else \expandafter \@secondoftwo
			\fi
		}%
		\providecommand \natexlab [1]{#1}%
		\providecommand \enquote  [1]{``#1''}%
		\providecommand \bibnamefont  [1]{#1}%
		\providecommand \bibfnamefont [1]{#1}%
		\providecommand \citenamefont [1]{#1}%
		\providecommand \href@noop [0]{\@secondoftwo}%
		\providecommand \href [0]{\begingroup \@sanitize@url \@href}%
		\providecommand \@href[1]{\@@startlink{#1}\@@href}%
		\providecommand \@@href[1]{\endgroup#1\@@endlink}%
		\providecommand \@sanitize@url [0]{\catcode `\\12\catcode `\$12\catcode
			`\&12\catcode `\#12\catcode `\^12\catcode `\_12\catcode `\%12\relax}%
		\providecommand \@@startlink[1]{}%
		\providecommand \@@endlink[0]{}%
		\providecommand \url  [0]{\begingroup\@sanitize@url \@url }%
		\providecommand \@url [1]{\endgroup\@href {#1}{\urlprefix }}%
		\providecommand \urlprefix  [0]{URL }%
		\providecommand \Eprint [0]{\href }%
		\providecommand \doibase [0]{http://dx.doi.org/}%
		\providecommand \selectlanguage [0]{\@gobble}%
		\providecommand \bibinfo  [0]{\@secondoftwo}%
		\providecommand \bibfield  [0]{\@secondoftwo}%
		\providecommand \translation [1]{[#1]}%
		\providecommand \BibitemOpen [0]{}%
		\providecommand \bibitemStop [0]{}%
		\providecommand \bibitemNoStop [0]{.\EOS\space}%
		\providecommand \EOS [0]{\spacefactor3000\relax}%
		\providecommand \BibitemShut  [1]{\csname bibitem#1\endcsname}%
		\let\auto@bib@innerbib\@empty
		\bibitem [{\citenamefont {Anderson}(1972)}]{anderson1972more}%
		\BibitemOpen
		\bibfield  {author} {\bibinfo {author} {\bibfnamefont {P.~W.}\ \bibnamefont
				{Anderson}},\ }\href@noop {} {\bibfield  {journal} {\bibinfo  {journal}
				{Science}\ }\textbf {\bibinfo {volume} {177}},\ \bibinfo {pages} {393}
			(\bibinfo {year} {1972})}\BibitemShut {NoStop}%
		\bibitem [{\citenamefont {Fisher}\ \emph {et~al.}(1989)\citenamefont {Fisher},
			\citenamefont {Weichman}, \citenamefont {Grinstein},\ and\ \citenamefont
			{Fisher}}]{fisher1989boson}%
		\BibitemOpen
		\bibfield  {author} {\bibinfo {author} {\bibfnamefont {M.~P.~A.}\ \bibnamefont
				{Fisher}}, \bibinfo {author} {\bibfnamefont {P.~B.}\ \bibnamefont
				{Weichman}}, \bibinfo {author} {\bibfnamefont {G.}~\bibnamefont {Grinstein}},
			\ and\ \bibinfo {author} {\bibfnamefont {D.~S.}\ \bibnamefont {Fisher}},\
		}\href@noop {} {\bibfield  {journal} {\bibinfo  {journal} {Physical Review
					B}\ }\textbf {\bibinfo {volume} {40}},\ \bibinfo {pages} {546} (\bibinfo
			{year} {1989})}\BibitemShut {NoStop}%
		\bibitem [{\citenamefont {Sachdev}(2007)}]{sachdev2007quantum}%
		\BibitemOpen
		\bibfield  {author} {\bibinfo {author} {\bibfnamefont {S.}~\bibnamefont
				{Sachdev}},\ }\href@noop {} {\bibfield  {journal} {\bibinfo  {journal}
				{Handbook of Magnetism and Advanced Magnetic Materials}\ } (\bibinfo {year}
			{2007})}\BibitemShut {NoStop}%
		\bibitem [{\citenamefont {Brush}(1967)}]{brush1967history}%
		\BibitemOpen
		\bibfield  {author} {\bibinfo {author} {\bibfnamefont {S.~G.}\ \bibnamefont
				{Brush}},\ }\href@noop {} {\bibfield  {journal} {\bibinfo  {journal} {Reviews
					of modern physics}\ }\textbf {\bibinfo {volume} {39}},\ \bibinfo {pages}
			{883} (\bibinfo {year} {1967})}\BibitemShut {NoStop}%
		\bibitem [{\citenamefont {Rigol}\ \emph {et~al.}(2008)\citenamefont {Rigol},
			\citenamefont {Dunjko},\ and\ \citenamefont
			{Olshanii}}]{rigol2008thermalization}%
		\BibitemOpen
		\bibfield  {author} {\bibinfo {author} {\bibfnamefont {M.}~\bibnamefont
				{Rigol}}, \bibinfo {author} {\bibfnamefont {V.}~\bibnamefont {Dunjko}}, \
			and\ \bibinfo {author} {\bibfnamefont {M.}~\bibnamefont {Olshanii}},\
		}\href@noop {} {\bibfield  {journal} {\bibinfo  {journal} {Nature}\ }\textbf
			{\bibinfo {volume} {452}},\ \bibinfo {pages} {854} (\bibinfo {year}
			{2008})}\BibitemShut {NoStop}%
		\bibitem [{\citenamefont {Kardar}(2007)}]{kardar2007statistical}%
		\BibitemOpen
		\bibfield  {author} {\bibinfo {author} {\bibfnamefont {M.}~\bibnamefont
				{Kardar}},\ }\href@noop {} {\emph {\bibinfo {title} {Statistical physics of
					particles}}}\ (\bibinfo  {publisher} {Cambridge University Press},\ \bibinfo
		{year} {2007})\BibitemShut {NoStop}%
		\bibitem [{Note1()}]{Note1}%
		\BibitemOpen
		\bibinfo {note} {Specifically, their expectation value in the quantum case,
			and average over space, time, or realizations of the initial conditions in
			the classical case.}\BibitemShut {Stop}%
		\bibitem [{Note2()}]{Note2}%
		\BibitemOpen
		\bibinfo {note} {Quoting from \cite {rigol2008thermalization}, ``In generic
			isolated systems, non-equilibrium dynamics is expected to result in
			thermalization [$\protect \dots $] However, it is not obvious what feature of
			many-body quantum mechanics makes quantum thermalization
			possible''.}\BibitemShut {Stop}%
		\bibitem [{\citenamefont {Canovi}\ \emph {et~al.}(2016)\citenamefont {Canovi},
			\citenamefont {Kollar},\ and\ \citenamefont
			{Eckstein}}]{canovi2016stroboscopic}%
		\BibitemOpen
		\bibfield  {author} {\bibinfo {author} {\bibfnamefont {E.}~\bibnamefont
				{Canovi}}, \bibinfo {author} {\bibfnamefont {M.}~\bibnamefont {Kollar}}, \
			and\ \bibinfo {author} {\bibfnamefont {M.}~\bibnamefont {Eckstein}},\
		}\href@noop {} {\bibfield  {journal} {\bibinfo  {journal} {Physical Review
					E}\ }\textbf {\bibinfo {volume} {93}},\ \bibinfo {pages} {012130} (\bibinfo
			{year} {2016})}\BibitemShut {NoStop}%
		\bibitem [{\citenamefont {Mori}\ \emph {et~al.}(2016)\citenamefont {Mori},
			\citenamefont {Kuwahara},\ and\ \citenamefont {Saito}}]{mori2016rigorous}%
		\BibitemOpen
		\bibfield  {author} {\bibinfo {author} {\bibfnamefont {T.}~\bibnamefont
				{Mori}}, \bibinfo {author} {\bibfnamefont {T.}~\bibnamefont {Kuwahara}}, \
			and\ \bibinfo {author} {\bibfnamefont {K.}~\bibnamefont {Saito}},\
		}\href@noop {} {\bibfield  {journal} {\bibinfo  {journal} {Physical review
					letters}\ }\textbf {\bibinfo {volume} {116}},\ \bibinfo {pages} {120401}
			(\bibinfo {year} {2016})}\BibitemShut {NoStop}%
		\bibitem [{\citenamefont {Abanin}\ \emph
			{et~al.}(2017{\natexlab{a}})\citenamefont {Abanin}, \citenamefont {De~Roeck},
			\citenamefont {Ho},\ and\ \citenamefont {Huveneers}}]{abanin2017effective}%
		\BibitemOpen
		\bibfield  {author} {\bibinfo {author} {\bibfnamefont {D.~A.}\ \bibnamefont
				{Abanin}}, \bibinfo {author} {\bibfnamefont {W.}~\bibnamefont {De~Roeck}},
			\bibinfo {author} {\bibfnamefont {W.~W.}\ \bibnamefont {Ho}}, \ and\ \bibinfo
			{author} {\bibfnamefont {F.}~\bibnamefont {Huveneers}},\ }\href@noop {}
		{\bibfield  {journal} {\bibinfo  {journal} {Physical Review B}\ }\textbf
			{\bibinfo {volume} {95}},\ \bibinfo {pages} {014112} (\bibinfo {year}
			{2017}{\natexlab{a}})}\BibitemShut {NoStop}%
		\bibitem [{\citenamefont {Weidinger}\ and\ \citenamefont
			{Knap}(2017)}]{weidinger2017floquet}%
		\BibitemOpen
		\bibfield  {author} {\bibinfo {author} {\bibfnamefont {S.~A.}\ \bibnamefont
				{Weidinger}}\ and\ \bibinfo {author} {\bibfnamefont {M.}~\bibnamefont
				{Knap}},\ }\href@noop {} {\bibfield  {journal} {\bibinfo  {journal}
				{Scientific Reports}\ }\textbf {\bibinfo {volume} {7}},\ \bibinfo {pages} {1}
			(\bibinfo {year} {2017})}\BibitemShut {NoStop}%
		\bibitem [{\citenamefont {Abanin}\ \emph
			{et~al.}(2017{\natexlab{b}})\citenamefont {Abanin}, \citenamefont {De~Roeck},
			\citenamefont {Ho},\ and\ \citenamefont {Huveneers}}]{abanin2017rigorous}%
		\BibitemOpen
		\bibfield  {author} {\bibinfo {author} {\bibfnamefont {D.}~\bibnamefont
				{Abanin}}, \bibinfo {author} {\bibfnamefont {W.}~\bibnamefont {De~Roeck}},
			\bibinfo {author} {\bibfnamefont {W.~W.}\ \bibnamefont {Ho}}, \ and\ \bibinfo
			{author} {\bibfnamefont {F.}~\bibnamefont {Huveneers}},\ }\href@noop {}
		{\bibfield  {journal} {\bibinfo  {journal} {Communications in Mathematical
					Physics}\ }\textbf {\bibinfo {volume} {354}},\ \bibinfo {pages} {809}
			(\bibinfo {year} {2017}{\natexlab{b}})}\BibitemShut {NoStop}%
		\bibitem [{\citenamefont {Mallayya}\ \emph {et~al.}(2019)\citenamefont
			{Mallayya}, \citenamefont {Rigol},\ and\ \citenamefont
			{De~Roeck}}]{mallayya2019prethermalization}%
		\BibitemOpen
		\bibfield  {author} {\bibinfo {author} {\bibfnamefont {K.}~\bibnamefont
				{Mallayya}}, \bibinfo {author} {\bibfnamefont {M.}~\bibnamefont {Rigol}}, \
			and\ \bibinfo {author} {\bibfnamefont {W.}~\bibnamefont {De~Roeck}},\
		}\href@noop {} {\bibfield  {journal} {\bibinfo  {journal} {Physical Review
					X}\ }\textbf {\bibinfo {volume} {9}},\ \bibinfo {pages} {021027} (\bibinfo
			{year} {2019})}\BibitemShut {NoStop}%
		\bibitem [{\citenamefont {Rajak}\ \emph {et~al.}(2018)\citenamefont {Rajak},
			\citenamefont {Citro},\ and\ \citenamefont
			{Dalla~Torre}}]{rajak2018stability}%
		\BibitemOpen
		\bibfield  {author} {\bibinfo {author} {\bibfnamefont {A.}~\bibnamefont
				{Rajak}}, \bibinfo {author} {\bibfnamefont {R.}~\bibnamefont {Citro}}, \ and\
			\bibinfo {author} {\bibfnamefont {E.~G.}\ \bibnamefont {Dalla~Torre}},\
		}\href@noop {} {\bibfield  {journal} {\bibinfo  {journal} {Journal of Physics
					A: Mathematical and Theoretical}\ }\textbf {\bibinfo {volume} {51}},\
			\bibinfo {pages} {465001} (\bibinfo {year} {2018})}\BibitemShut {NoStop}%
		\bibitem [{\citenamefont {Mori}(2018)}]{mori2018floquet}%
		\BibitemOpen
		\bibfield  {author} {\bibinfo {author} {\bibfnamefont {T.}~\bibnamefont
				{Mori}},\ }\href@noop {} {\bibfield  {journal} {\bibinfo  {journal} {Physical
					Review B}\ }\textbf {\bibinfo {volume} {98}},\ \bibinfo {pages} {104303}
			(\bibinfo {year} {2018})}\BibitemShut {NoStop}%
		\bibitem [{\citenamefont {Rajak}\ \emph {et~al.}(2019)\citenamefont {Rajak},
			\citenamefont {Dana},\ and\ \citenamefont
			{Dalla~Torre}}]{rajak2019characterizations}%
		\BibitemOpen
		\bibfield  {author} {\bibinfo {author} {\bibfnamefont {A.}~\bibnamefont
				{Rajak}}, \bibinfo {author} {\bibfnamefont {I.}~\bibnamefont {Dana}}, \ and\
			\bibinfo {author} {\bibfnamefont {E.~G.}\ \bibnamefont {Dalla~Torre}},\
		}\href@noop {} {\bibfield  {journal} {\bibinfo  {journal} {Physical Review
					B}\ }\textbf {\bibinfo {volume} {100}},\ \bibinfo {pages} {100302(R)} (\bibinfo
			{year} {2019})}\BibitemShut {NoStop}%
		\bibitem [{\citenamefont {Howell}\ \emph {et~al.}(2019)\citenamefont {Howell},
			\citenamefont {Weinberg}, \citenamefont {Sels}, \citenamefont {Polkovnikov},\
			and\ \citenamefont {Bukov}}]{howell2019asymptotic}%
		\BibitemOpen
		\bibfield  {author} {\bibinfo {author} {\bibfnamefont {O.}~\bibnamefont
				{Howell}}, \bibinfo {author} {\bibfnamefont {P.}~\bibnamefont {Weinberg}},
			\bibinfo {author} {\bibfnamefont {D.}~\bibnamefont {Sels}}, \bibinfo {author}
			{\bibfnamefont {A.}~\bibnamefont {Polkovnikov}}, \ and\ \bibinfo {author}
			{\bibfnamefont {M.}~\bibnamefont {Bukov}},\ }\href@noop {} {\bibfield
			{journal} {\bibinfo  {journal} {Physical Review Letters}\ }\textbf {\bibinfo
				{volume} {122}},\ \bibinfo {pages} {010602} (\bibinfo {year}
			{2019})}\BibitemShut {NoStop}%
		\bibitem [{\citenamefont {Else}\ \emph {et~al.}(2017)\citenamefont {Else},
			\citenamefont {Bauer},\ and\ \citenamefont {Nayak}}]{else2017prethermal}%
		\BibitemOpen
		\bibfield  {author} {\bibinfo {author} {\bibfnamefont {D.~V.}\ \bibnamefont
				{Else}}, \bibinfo {author} {\bibfnamefont {B.}~\bibnamefont {Bauer}}, \ and\
			\bibinfo {author} {\bibfnamefont {C.}~\bibnamefont {Nayak}},\ }\href@noop {}
		{\bibfield  {journal} {\bibinfo  {journal} {Physical Review X}\ }\textbf
			{\bibinfo {volume} {7}},\ \bibinfo {pages} {011026} (\bibinfo {year}
			{2017})}\BibitemShut {NoStop}%
		\bibitem [{\citenamefont {Machado}\ \emph {et~al.}(2020)\citenamefont
			{Machado}, \citenamefont {Else}, \citenamefont {Kahanamoku-Meyer},
			\citenamefont {Nayak},\ and\ \citenamefont {Yao}}]{machado2020long}%
		\BibitemOpen
		\bibfield  {author} {\bibinfo {author} {\bibfnamefont {F.}~\bibnamefont
				{Machado}}, \bibinfo {author} {\bibfnamefont {D.~V.}\ \bibnamefont {Else}},
			\bibinfo {author} {\bibfnamefont {G.~D.}\ \bibnamefont {Kahanamoku-Meyer}},
			\bibinfo {author} {\bibfnamefont {C.}~\bibnamefont {Nayak}}, \ and\ \bibinfo
			{author} {\bibfnamefont {N.~Y.}\ \bibnamefont {Yao}},\ }\href@noop {}
		{\bibfield  {journal} {\bibinfo  {journal} {Physical Review X}\ }\textbf
			{\bibinfo {volume} {10}},\ \bibinfo {pages} {011043} (\bibinfo {year}
			{2020})}\BibitemShut {NoStop}%
		\bibitem [{\citenamefont {Luitz}\ \emph {et~al.}(2020)\citenamefont {Luitz},
			\citenamefont {Moessner}, \citenamefont {Sondhi},\ and\ \citenamefont
			{Khemani}}]{luitz2020prethermalization}%
		\BibitemOpen
		\bibfield  {author} {\bibinfo {author} {\bibfnamefont {D.~J.}\ \bibnamefont
				{Luitz}}, \bibinfo {author} {\bibfnamefont {R.}~\bibnamefont {Moessner}},
			\bibinfo {author} {\bibfnamefont {S.~L.}~\bibnamefont {Sondhi}}, \ and\ \bibinfo
			{author} {\bibfnamefont {V.}~\bibnamefont {Khemani}},\ }\href@noop {}
		{\bibfield  {journal} {\bibinfo  {journal} {Physical Review X}\ }\textbf
			{\bibinfo {volume} {10}},\ \bibinfo {pages} {021046} (\bibinfo {year}
			{2020})}\BibitemShut {NoStop}%
		\bibitem [{\citenamefont {Zhao}\ \emph {et~al.}(2021)\citenamefont {Zhao},
			\citenamefont {Mintert}, \citenamefont {Moessner},\ and\ \citenamefont
			{Knolle}}]{zhao2021random}%
		\BibitemOpen
		\bibfield  {author} {\bibinfo {author} {\bibfnamefont {H.}~\bibnamefont
				{Zhao}}, \bibinfo {author} {\bibfnamefont {F.}~\bibnamefont {Mintert}},
			\bibinfo {author} {\bibfnamefont {R.}~\bibnamefont {Moessner}}, \ and\
			\bibinfo {author} {\bibfnamefont {J.}~\bibnamefont {Knolle}},\ }\href@noop {}
		{\bibfield  {journal} {\bibinfo  {journal} {Physical Review Letters}\
			}\textbf {\bibinfo {volume} {126}},\ \bibinfo {pages} {040601} (\bibinfo
			{year} {2021})}\BibitemShut {NoStop}%
		\bibitem [{\citenamefont {Sacha}(2015)}]{sacha2015modeling}%
		\BibitemOpen
		\bibfield  {author} {\bibinfo {author} {\bibfnamefont {K.}~\bibnamefont
				{Sacha}},\ }\href@noop {} {\bibfield  {journal} {\bibinfo  {journal}
				{Physical Review A}\ }\textbf {\bibinfo {volume} {91}},\ \bibinfo {pages}
			{033617} (\bibinfo {year} {2015})}\BibitemShut {NoStop}%
		\bibitem [{\citenamefont {Khemani}\ \emph {et~al.}(2016)\citenamefont
			{Khemani}, \citenamefont {Lazarides}, \citenamefont {Moessner},\ and\
			\citenamefont {Sondhi}}]{khemani2016phase}%
		\BibitemOpen
		\bibfield  {author} {\bibinfo {author} {\bibfnamefont {V.}~\bibnamefont
				{Khemani}}, \bibinfo {author} {\bibfnamefont {A.}~\bibnamefont {Lazarides}},
			\bibinfo {author} {\bibfnamefont {R.}~\bibnamefont {Moessner}}, \ and\
			\bibinfo {author} {\bibfnamefont {S.~L.}\ \bibnamefont {Sondhi}},\
		}\href@noop {} {\bibfield  {journal} {\bibinfo  {journal} {Physical Review
					Letters}\ }\textbf {\bibinfo {volume} {116}},\ \bibinfo {pages} {250401}
			(\bibinfo {year} {2016})}\BibitemShut {NoStop}%
		\bibitem [{\citenamefont {Else}\ \emph {et~al.}(2016)\citenamefont {Else},
			\citenamefont {Bauer},\ and\ \citenamefont {Nayak}}]{else2016floquet}%
		\BibitemOpen
		\bibfield  {author} {\bibinfo {author} {\bibfnamefont {D.~V.}\ \bibnamefont
				{Else}}, \bibinfo {author} {\bibfnamefont {B.}~\bibnamefont {Bauer}}, \ and\
			\bibinfo {author} {\bibfnamefont {C.}~\bibnamefont {Nayak}},\ }\href@noop {}
		{\bibfield  {journal} {\bibinfo  {journal} {Physical Review Letters}\
			}\textbf {\bibinfo {volume} {117}},\ \bibinfo {pages} {090402} (\bibinfo
			{year} {2016})}\BibitemShut {NoStop}%
		\bibitem [{\citenamefont {Yao}\ \emph {et~al.}(2017)\citenamefont {Yao},
			\citenamefont {Potter}, \citenamefont {Potirniche},\ and\ \citenamefont
			{Vishwanath}}]{yao2017discrete}%
		\BibitemOpen
		\bibfield  {author} {\bibinfo {author} {\bibfnamefont {N.~Y.}\ \bibnamefont
				{Yao}}, \bibinfo {author} {\bibfnamefont {A.~C.}\ \bibnamefont {Potter}},
			\bibinfo {author} {\bibfnamefont {I.-D.}\ \bibnamefont {Potirniche}}, \ and\
			\bibinfo {author} {\bibfnamefont {A.}~\bibnamefont {Vishwanath}},\
		}\href@noop {} {\bibfield  {journal} {\bibinfo  {journal} {Physical Review
					Letters}\ }\textbf {\bibinfo {volume} {118}},\ \bibinfo {pages} {030401}
			(\bibinfo {year} {2017})}\BibitemShut {NoStop}%
		\bibitem [{\citenamefont {Moessner}\ and\ \citenamefont
			{Sondhi}(2017)}]{moessner2017equilibration}%
		\BibitemOpen
		\bibfield  {author} {\bibinfo {author} {\bibfnamefont {R.}~\bibnamefont
				{Moessner}}\ and\ \bibinfo {author} {\bibfnamefont {S.~L.}\ \bibnamefont
				{Sondhi}},\ }\href@noop {} {\bibfield  {journal} {\bibinfo  {journal} {Nature
					Physics}\ }\textbf {\bibinfo {volume} {13}},\ \bibinfo {pages} {424}
			(\bibinfo {year} {2017})}\BibitemShut {NoStop}%
		\bibitem [{\citenamefont {von Keyserlingk}\ \emph {et~al.}(2016)\citenamefont
			{von Keyserlingk}, \citenamefont {Khemani},\ and\ \citenamefont
			{Sondhi}}]{von2016absolute}%
		\BibitemOpen
		\bibfield  {author} {\bibinfo {author} {\bibfnamefont {C.~W.}\ \bibnamefont
				{von Keyserlingk}}, \bibinfo {author} {\bibfnamefont {V.}~\bibnamefont
				{Khemani}}, \ and\ \bibinfo {author} {\bibfnamefont {S.~L.}\ \bibnamefont
				{Sondhi}},\ }\href@noop {} {\bibfield  {journal} {\bibinfo  {journal}
				{Physical Review B}\ }\textbf {\bibinfo {volume} {94}},\ \bibinfo {pages}
			{085112} (\bibinfo {year} {2016})}\BibitemShut {NoStop}%
		\bibitem [{\citenamefont {Gong}\ \emph {et~al.}(2018)\citenamefont {Gong},
			\citenamefont {Hamazaki},\ and\ \citenamefont {Ueda}}]{gong2018discrete}%
		\BibitemOpen
		\bibfield  {author} {\bibinfo {author} {\bibfnamefont {Z.}~\bibnamefont
				{Gong}}, \bibinfo {author} {\bibfnamefont {R.}~\bibnamefont {Hamazaki}}, \
			and\ \bibinfo {author} {\bibfnamefont {M.}~\bibnamefont {Ueda}},\ }\href@noop
		{} {\bibfield  {journal} {\bibinfo  {journal} {Physical review letters}\
			}\textbf {\bibinfo {volume} {120}},\ \bibinfo {pages} {040404} (\bibinfo
			{year} {2018})}\BibitemShut {NoStop}%
		\bibitem [{\citenamefont {Giergiel}\ \emph {et~al.}(2019)\citenamefont
			{Giergiel}, \citenamefont {Kuro{\'s}},\ and\ \citenamefont
			{Sacha}}]{giergiel2019discrete}%
		\BibitemOpen
		\bibfield  {author} {\bibinfo {author} {\bibfnamefont {K.}~\bibnamefont
				{Giergiel}}, \bibinfo {author} {\bibfnamefont {A.}~\bibnamefont {Kuro{\'s}}},
			\ and\ \bibinfo {author} {\bibfnamefont {K.}~\bibnamefont {Sacha}},\
		}\href@noop {} {\bibfield  {journal} {\bibinfo  {journal} {Physical Review
					B}\ }\textbf {\bibinfo {volume} {99}},\ \bibinfo {pages} {220303(R)} (\bibinfo
			{year} {2019})}\BibitemShut {NoStop}%
		\bibitem [{\citenamefont {Matus}\ and\ \citenamefont
			{Sacha}(2019)}]{matus2019fractional}%
		\BibitemOpen
		\bibfield  {author} {\bibinfo {author} {\bibfnamefont {P.}~\bibnamefont
				{Matus}}\ and\ \bibinfo {author} {\bibfnamefont {K.}~\bibnamefont {Sacha}},\
		}\href@noop {} {\bibfield  {journal} {\bibinfo  {journal} {Physical Review
					A}\ }\textbf {\bibinfo {volume} {99}},\ \bibinfo {pages} {033626} (\bibinfo
			{year} {2019})}\BibitemShut {NoStop}%
		\bibitem [{\citenamefont {Gambetta}\ \emph
			{et~al.}(2019{\natexlab{a}})\citenamefont {Gambetta}, \citenamefont
			{Carollo}, \citenamefont {Marcuzzi}, \citenamefont {Garrahan},\ and\
			\citenamefont {Lesanovsky}}]{gambetta2019discrete}%
		\BibitemOpen
		\bibfield  {author} {\bibinfo {author} {\bibfnamefont {F.~M.}~\bibnamefont
				{Gambetta}}, \bibinfo {author} {\bibfnamefont {F.}~\bibnamefont {Carollo}},
			\bibinfo {author} {\bibfnamefont {M.}~\bibnamefont {Marcuzzi}}, \bibinfo
			{author} {\bibfnamefont {J.~P.}~\bibnamefont {Garrahan}}, \ and\ \bibinfo
			{author} {\bibfnamefont {I.}~\bibnamefont {Lesanovsky}},\ }\href@noop {}
		{\bibfield  {journal} {\bibinfo  {journal} {Physical review letters}\
			}\textbf {\bibinfo {volume} {122}},\ \bibinfo {pages} {015701} (\bibinfo
			{year} {2019}{\natexlab{a}})}\BibitemShut {NoStop}%
		\bibitem [{\citenamefont {Gambetta}\ \emph
			{et~al.}(2019{\natexlab{b}})\citenamefont {Gambetta}, \citenamefont
			{Carollo}, \citenamefont {Lazarides}, \citenamefont {Lesanovsky},\ and\
			\citenamefont {Garrahan}}]{gambetta2019classical}%
		\BibitemOpen
		\bibfield  {author} {\bibinfo {author} {\bibfnamefont {F.~M.}~\bibnamefont
				{Gambetta}}, \bibinfo {author} {\bibfnamefont {F.}~\bibnamefont {Carollo}},
			\bibinfo {author} {\bibfnamefont {A.}~\bibnamefont {Lazarides}}, \bibinfo
			{author} {\bibfnamefont {I.}~\bibnamefont {Lesanovsky}}, \ and\ \bibinfo
			{author} {\bibfnamefont {J.~P.}~\bibnamefont {Garrahan}},\ }\href@noop {}
		{\bibfield  {journal} {\bibinfo  {journal} {Physical Review E}\ }\textbf
			{\bibinfo {volume} {100}},\ \bibinfo {pages} {060105(R)} (\bibinfo {year}
			{2019}{\natexlab{b}})}\BibitemShut {NoStop}%
		\bibitem [{\citenamefont {Zhu}\ \emph {et~al.}(2019)\citenamefont {Zhu},
			\citenamefont {Marino}, \citenamefont {Yao}, \citenamefont {Lukin},\ and\
			\citenamefont {Demler}}]{zhu2019dicke}%
		\BibitemOpen
		\bibfield  {author} {\bibinfo {author} {\bibfnamefont {B.}~\bibnamefont
				{Zhu}}, \bibinfo {author} {\bibfnamefont {J.}~\bibnamefont {Marino}},
			\bibinfo {author} {\bibfnamefont {N.~Y.}\ \bibnamefont {Yao}}, \bibinfo
			{author} {\bibfnamefont {M.~D.}\ \bibnamefont {Lukin}}, \ and\ \bibinfo
			{author} {\bibfnamefont {E.~A.}\ \bibnamefont {Demler}},\ }\href@noop {}
		{\bibfield  {journal} {\bibinfo  {journal} {New Journal of Physics}\ }\textbf
			{\bibinfo {volume} {21}},\ \bibinfo {pages} {073028} (\bibinfo {year}
			{2019})}\BibitemShut {NoStop}%
		\bibitem [{\citenamefont {Ke{\ss}ler}\ \emph {et~al.}(2019)\citenamefont
			{Ke{\ss}ler}, \citenamefont {Cosme}, \citenamefont {Hemmerling},
			\citenamefont {Mathey},\ and\ \citenamefont
			{Hemmerich}}]{kessler2019emergent}%
		\BibitemOpen
		\bibfield  {author} {\bibinfo {author} {\bibfnamefont {H.}~\bibnamefont
				{Ke{\ss}ler}}, \bibinfo {author} {\bibfnamefont {J.~G.}\ \bibnamefont
				{Cosme}}, \bibinfo {author} {\bibfnamefont {M.}~\bibnamefont {Hemmerling}},
			\bibinfo {author} {\bibfnamefont {L.}~\bibnamefont {Mathey}}, \ and\ \bibinfo
			{author} {\bibfnamefont {A.}~\bibnamefont {Hemmerich}},\ }\href@noop {}
		{\bibfield  {journal} {\bibinfo  {journal} {Physical Review A}\ }\textbf
			{\bibinfo {volume} {99}},\ \bibinfo {pages} {053605} (\bibinfo {year}
			{2019})}\BibitemShut {NoStop}%
		\bibitem [{\citenamefont {Pizzi}\ \emph {et~al.}(2019)\citenamefont {Pizzi},
			\citenamefont {Knolle},\ and\ \citenamefont {Nunnenkamp}}]{pizzi2019period}%
		\BibitemOpen
		\bibfield  {author} {\bibinfo {author} {\bibfnamefont {A.}~\bibnamefont
				{Pizzi}}, \bibinfo {author} {\bibfnamefont {J.}~\bibnamefont {Knolle}}, \
			and\ \bibinfo {author} {\bibfnamefont {A.}~\bibnamefont {Nunnenkamp}},\
		}\href@noop {} {\bibfield  {journal} {\bibinfo  {journal} {Phys. Rev. Lett.}\
			}\textbf {\bibinfo {volume} {123}},\ \bibinfo {pages} {150601} (\bibinfo
			{year} {2019})}\BibitemShut {NoStop}%
		\bibitem [{\citenamefont {Yao}\ \emph {et~al.}(2020)\citenamefont {Yao},
			\citenamefont {Nayak}, \citenamefont {Balents},\ and\ \citenamefont
			{Zaletel}}]{yao2020classical}%
		\BibitemOpen
		\bibfield  {author} {\bibinfo {author} {\bibfnamefont {N.~Y.}\ \bibnamefont
				{Yao}}, \bibinfo {author} {\bibfnamefont {C.}~\bibnamefont {Nayak}}, \bibinfo
			{author} {\bibfnamefont {L.}~\bibnamefont {Balents}}, \ and\ \bibinfo
			{author} {\bibfnamefont {M.~P.}\ \bibnamefont {Zaletel}},\ }\href@noop {}
		{\bibfield  {journal} {\bibinfo  {journal} {Nature Physics}\ }\textbf
			{\bibinfo {volume} {16}},\ \bibinfo {pages} {438} (\bibinfo {year}
			{2020})}\BibitemShut {NoStop}%
		\bibitem [{\citenamefont {Pizzi}\ \emph
			{et~al.}(2021{\natexlab{a}})\citenamefont {Pizzi}, \citenamefont
			{Nunnenkamp},\ and\ \citenamefont {Knolle}}]{pizzi2021bistability}%
		\BibitemOpen
		\bibfield  {author} {\bibinfo {author} {\bibfnamefont {A.}~\bibnamefont
				{Pizzi}}, \bibinfo {author} {\bibfnamefont {A.}~\bibnamefont {Nunnenkamp}}, \
			and\ \bibinfo {author} {\bibfnamefont {J.}~\bibnamefont {Knolle}},\
		}\href@noop {} {\bibfield  {journal} {\bibinfo  {journal} {Nature
					Communications}\ }\textbf {\bibinfo {volume} {12}},\ \bibinfo {pages} {1061}
			(\bibinfo {year} {2021}{\natexlab{a}})}\BibitemShut {NoStop}%
		\bibitem [{\citenamefont {Malz}\ \emph {et~al.}(2021)\citenamefont {Malz},
			\citenamefont {Pizzi}, \citenamefont {Nunnenkamp},\ and\ \citenamefont
			{Knolle}}]{malz2021seasonal}%
		\BibitemOpen
		\bibfield  {author} {\bibinfo {author} {\bibfnamefont {D.}~\bibnamefont
				{Malz}}, \bibinfo {author} {\bibfnamefont {A.}~\bibnamefont {Pizzi}},
			\bibinfo {author} {\bibfnamefont {A.}~\bibnamefont {Nunnenkamp}}, \ and\
			\bibinfo {author} {\bibfnamefont {J.}~\bibnamefont {Knolle}},\ }\href@noop {}
		{\bibfield  {journal} {\bibinfo  {journal} {Physical Review Research}\
			}\textbf {\bibinfo {volume} {3}},\ \bibinfo {pages} {013124} (\bibinfo {year}
			{2021})}\BibitemShut {NoStop}%
		\bibitem [{\citenamefont {Choi}\ \emph {et~al.}(2017)\citenamefont {Choi},
			\citenamefont {Choi}, \citenamefont {Landig}, \citenamefont {Kucsko},
			\citenamefont {Zhou}, \citenamefont {Isoya}, \citenamefont {Jelezko},
			\citenamefont {Onoda}, \citenamefont {Sumiya}, \citenamefont {Khemani} \emph
			{et~al.}}]{choi2017observation}%
		\BibitemOpen
		\bibfield  {author} {\bibinfo {author} {\bibfnamefont {S.}~\bibnamefont
				{Choi}}, \bibinfo {author} {\bibfnamefont {J.}~\bibnamefont {Choi}}, \bibinfo
			{author} {\bibfnamefont {R.}~\bibnamefont {Landig}}, \bibinfo {author}
			{\bibfnamefont {G.}~\bibnamefont {Kucsko}}, \bibinfo {author} {\bibfnamefont
				{H.}~\bibnamefont {Zhou}}, \bibinfo {author} {\bibfnamefont {J.}~\bibnamefont
				{Isoya}}, \bibinfo {author} {\bibfnamefont {F.}~\bibnamefont {Jelezko}},
			\bibinfo {author} {\bibfnamefont {S.}~\bibnamefont {Onoda}}, \bibinfo
			{author} {\bibfnamefont {H.}~\bibnamefont {Sumiya}}, \bibinfo {author}
			{\bibfnamefont {V.}~\bibnamefont {Khemani}},  \emph {et~al.},\ }\href@noop {}
		{\bibfield  {journal} {\bibinfo  {journal} {Nature}\ }\textbf {\bibinfo
				{volume} {543}},\ \bibinfo {pages} {221} (\bibinfo {year}
			{2017})}\BibitemShut {NoStop}%
		\bibitem [{\citenamefont {Zhang}\ \emph {et~al.}(2017)\citenamefont {Zhang},
			\citenamefont {Hess}, \citenamefont {Kyprianidis}, \citenamefont {Becker},
			\citenamefont {Lee}, \citenamefont {Smith}, \citenamefont {Pagano},
			\citenamefont {Potirniche}, \citenamefont {Potter}, \citenamefont
			{Vishwanath} \emph {et~al.}}]{zhang2017observation}%
		\BibitemOpen
		\bibfield  {author} {\bibinfo {author} {\bibfnamefont {J.}~\bibnamefont
				{Zhang}}, \bibinfo {author} {\bibfnamefont {P.}~\bibnamefont {Hess}},
			\bibinfo {author} {\bibfnamefont {A.}~\bibnamefont {Kyprianidis}}, \bibinfo
			{author} {\bibfnamefont {P.}~\bibnamefont {Becker}}, \bibinfo {author}
			{\bibfnamefont {A.}~\bibnamefont {Lee}}, \bibinfo {author} {\bibfnamefont
				{J.}~\bibnamefont {Smith}}, \bibinfo {author} {\bibfnamefont
				{G.}~\bibnamefont {Pagano}}, \bibinfo {author} {\bibfnamefont {I.-D.}\
				\bibnamefont {Potirniche}}, \bibinfo {author} {\bibfnamefont {A.~C.}\
				\bibnamefont {Potter}}, \bibinfo {author} {\bibfnamefont {A.}~\bibnamefont
				{Vishwanath}},  \emph {et~al.},\ }\href@noop {} {\bibfield  {journal}
			{\bibinfo  {journal} {Nature}\ }\textbf {\bibinfo {volume} {543}},\ \bibinfo
			{pages} {217} (\bibinfo {year} {2017})}\BibitemShut {NoStop}%
		\bibitem [{\citenamefont {Rovny}\ \emph {et~al.}(2018)\citenamefont {Rovny},
			\citenamefont {Blum},\ and\ \citenamefont {Barrett}}]{rovny2018observation}%
		\BibitemOpen
		\bibfield  {author} {\bibinfo {author} {\bibfnamefont {J.}~\bibnamefont
				{Rovny}}, \bibinfo {author} {\bibfnamefont {R.~L.}\ \bibnamefont {Blum}}, \
			and\ \bibinfo {author} {\bibfnamefont {S.~E.}\ \bibnamefont {Barrett}},\
		}\href@noop {} {\bibfield  {journal} {\bibinfo  {journal} {Physical Review
					Letters}\ }\textbf {\bibinfo {volume} {120}},\ \bibinfo {pages} {180603}
			(\bibinfo {year} {2018})}\BibitemShut {NoStop}%
		\bibitem [{\citenamefont {Pizzi}\ \emph {et~al.}(2021)\citenamefont {Pizzi},
			\citenamefont {Nunnenkamp}, and\ \citenamefont {Knolle}}]{pizzi2021classicala}%
		\BibitemOpen
		\bibfield  {author} {\bibinfo {author} {\bibfnamefont {A.}~\bibnamefont
				{Pizzi}}, \bibinfo {author} {\bibfnamefont {A.}~\bibnamefont {Nunnenkamp}}, \ and\
			\bibinfo {author} {\bibfnamefont {J.}~\bibnamefont {Knolle}},\ }\href@noop {}
		{\bibfield  {journal} {\bibinfo  {journal} {Physical Review Letters}\
			}\textbf {\bibinfo {volume} {127}},\ \bibinfo {pages} {140602} (\bibinfo
			{year} {2021})}\BibitemShut {NoStop}%
		\bibitem [{\citenamefont {Bingtian}\ \emph {et~al.}(2021)\citenamefont {Bingtian},
			\citenamefont {Machado}, and\ \citenamefont {Yao}}]{ye2021classical}%
		\BibitemOpen
		\bibfield  {author} {\bibinfo {author} {\bibfnamefont {Y.}~\bibnamefont
				{Bingtian}}, \bibinfo {author} {\bibfnamefont {F.}~\bibnamefont {Machado}}, \ and\
			\bibinfo {author} {\bibfnamefont {N.~Y.}~\bibnamefont {Yao}},\ }\href@noop {}
		{\bibfield  {journal} {\bibinfo  {journal} {Physical Review Letters}\
			}\textbf {\bibinfo {volume} {127}},\ \bibinfo {pages} {140603} (\bibinfo
			{year} {2021})}\BibitemShut {NoStop}%
		\bibitem [{\citenamefont {Pizzi}\ \emph
			{et~al.}(2021{\natexlab{c}})\citenamefont {Pizzi}, \citenamefont {Knolle},\
			and\ \citenamefont {Nunnenkamp}}]{pizzi2021higher}%
		\BibitemOpen
		\bibfield  {author} {\bibinfo {author} {\bibfnamefont {A.}~\bibnamefont
				{Pizzi}}, \bibinfo {author} {\bibfnamefont {J.}~\bibnamefont {Knolle}}, \
			and\ \bibinfo {author} {\bibfnamefont {A.}~\bibnamefont {Nunnenkamp}},\
		}\href@noop {} {\bibfield  {journal} {\bibinfo  {journal} {Nature
					Communications}\ }\textbf {\bibinfo {volume} {12}},\ \bibinfo {pages} {2341}
			(\bibinfo {year} {2021}{\natexlab{c}})}\BibitemShut {NoStop}%
		\bibitem [{\citenamefont {Khasseh}\ \emph {et~al.}(2019)\citenamefont
			{Khasseh}, \citenamefont {Fazio}, \citenamefont {Ruffo},\ and\ \citenamefont
			{Russomanno}}]{khasseh2019many}%
		\BibitemOpen
		\bibfield  {author} {\bibinfo {author} {\bibfnamefont {R.}~\bibnamefont
				{Khasseh}}, \bibinfo {author} {\bibfnamefont {R.}~\bibnamefont {Fazio}},
			\bibinfo {author} {\bibfnamefont {S.}~\bibnamefont {Ruffo}}, \ and\ \bibinfo
			{author} {\bibfnamefont {A.}~\bibnamefont {Russomanno}},\ }\href@noop {}
		{\bibfield  {journal} {\bibinfo  {journal} {Physical Review Letters}\
			}\textbf {\bibinfo {volume} {123}},\ \bibinfo {pages} {184301} (\bibinfo
			{year} {2019})}\BibitemShut {NoStop}%
		\bibitem [{\citenamefont {Bilitewski}\ \emph {et~al.}(2018)\citenamefont
			{Bilitewski}, \citenamefont {Bhattacharjee},\ and\ \citenamefont
			{Moessner}}]{bilitewski2018temperature}%
		\BibitemOpen
		\bibfield  {author} {\bibinfo {author} {\bibfnamefont {T.}~\bibnamefont
				{Bilitewski}}, \bibinfo {author} {\bibfnamefont {S.}~\bibnamefont
				{Bhattacharjee}}, \ and\ \bibinfo {author} {\bibfnamefont {R.}~\bibnamefont
				{Moessner}},\ }\href@noop {} {\bibfield  {journal} {\bibinfo  {journal}
				{Physical Review Letters}\ }\textbf {\bibinfo {volume} {121}},\ \bibinfo
			{pages} {250602} (\bibinfo {year} {2018})}\BibitemShut {NoStop}%
		\bibitem [{\citenamefont {Bilitewski}\ \emph {et~al.}(2020)\citenamefont
			{Bilitewski}, \citenamefont {Bhattacharjee},\ and\ \citenamefont
			{Moessner}}]{bilitewski2020classical}%
		\BibitemOpen
		\bibfield  {author} {\bibinfo {author} {\bibfnamefont {T.}~\bibnamefont
				{Bilitewski}}, \bibinfo {author} {\bibfnamefont {S.}~\bibnamefont
				{Bhattacharjee}}, \ and\ \bibinfo {author} {\bibfnamefont {R.}~\bibnamefont
				{Moessner}},\ }\href@noop {} {\bibfield  {journal} {\bibinfo  {journal}
				{arXiv:2011.04700}\ } (\bibinfo {year} {2020})}\BibitemShut {NoStop}%
		\bibitem [{\citenamefont {Grundler}(2002)}]{grundler2002spintronics}%
		\BibitemOpen
		\bibfield  {author} {\bibinfo {author} {\bibfnamefont {D.}~\bibnamefont
				{Grundler}},\ }\href@noop {} {\bibfield  {journal} {\bibinfo  {journal}
				{Physics World}\ }\textbf {\bibinfo {volume} {15}},\ \bibinfo {pages} {39}
			(\bibinfo {year} {2002})}\BibitemShut {NoStop}%
		\bibitem [{\citenamefont {Kruglyak}\ \emph {et~al.}(2010)\citenamefont
			{Kruglyak}, \citenamefont {Demokritov},\ and\ \citenamefont
			{Grundler}}]{kruglyak2010magnonics}%
		\BibitemOpen
		\bibfield  {author} {\bibinfo {author} {\bibfnamefont {V.}~\bibnamefont
				{Kruglyak}}, \bibinfo {author} {\bibfnamefont {S.}~\bibnamefont
				{Demokritov}}, \ and\ \bibinfo {author} {\bibfnamefont {D.}~\bibnamefont
				{Grundler}},\ }\href@noop {} {\bibfield  {journal} {\bibinfo  {journal}
				{Journal of Physics D: Applied Physics}\ }\textbf {\bibinfo {volume} {43}},\
			\bibinfo {pages} {264001} (\bibinfo {year} {2010})}\BibitemShut {NoStop}%
		\bibitem [{Note3()}]{Note3}%
		\BibitemOpen
		\bibinfo {note} {Theories and models can be labelled as classical or quantum,
			but if both classical and quantum theories can shed light on prethermal DTCs,
			then a classification of prethermal DTCs as classical or quantum is perhaps
			not needed.}\BibitemShut {Stop}%
	\end{thebibliography}
\end{document}